\documentclass[showpacs,preprint]{revtex4}

\usepackage{graphicx}
\usepackage{dcolumn}
\usepackage{amsmath}
\usepackage[latin1]{inputenc}
\usepackage{graphicx}
\usepackage{amssymb}
\usepackage[colorlinks=true, citecolor=blue, urlcolor = blue, linkcolor= red, bookmarks=true]{hyperref}
\usepackage{float}
\usepackage{amsmath}
\usepackage{tensor}
\usepackage{amsfonts}
\usepackage{dcolumn}
\usepackage{hyperref}
\usepackage{subfigure}
\usepackage{pgfplots}
\usepackage{epstopdf}
\usepackage{booktabs}
\def \beq{\begin{equation}}
\def \eeq{\end{equation}}
\def \bse{\begin{subequations}}
\def \ese{\end{subequations}}
\def \bea{\begin{eqnarray}}
\def \eea{\end{eqnarray}}
\def \bem{\begin{displaymath}}
\def \eem{\end{displaymath}}
\def \bem{\begin{pmatrix}}
\def \eem{\end{pmatrix}}
\def \beb{\begin{bmatrix}}
\def \eeb{\end{bmatrix}}
\def \bc{\begin{center}}
\def \ec{\end{center}}

\def \nn{\nonumber}

\makeatletter
\def\btt#1{\texttt{\@backslashchar#1}}
\DeclareRobustCommand\bblash{\btt{\@backslashchar}} \makeatother

\makeatletter
\def\btt#1{\texttt{\@backslashchar#1}}
\DeclareRobustCommand\bblash{\btt{\@backslashchar}} \makeatother

\begin{document}

\title[]{ Rotating black hole in Rastall theory}

\author{Rahul Kumar$^{a}$}\email{rahul.phy3@gmail.com}

\author{Sushant~G.~Ghosh$^{a,\;b,\;c,}$} \email{sghosh2@jmi.ac.in, sgghosh@gmail.com}
\affiliation{$^{a}$ Centre for Theoretical Physics, Jamia Millia
Islamia, New Delhi 110025, India}
\affiliation{$^{b}$ Multidisciplinary Centre for Advanced Research and Studies (MCARS),
Jamia Millia Islamia, New Delhi 110025, India}
\affiliation{$^{c}$ Astrophysics and Cosmology
Research Unit, School of Mathematics, Statistics and Computer Science, University of
KwaZulu-Natal, Private Bag 54001, Durban 4000, South Africa}


\begin{abstract}
Rotating black hole solutions in theories of modified gravity are important as they offer an arena to test these
theories through astrophysical observation. The non-rotating black hole can be hardly tested since the black hole spin is
very important in any astrophysical process. We present rotating counterpart of a recently obtained spherically symmetric
exact black hole solution surrounded by perfect fluid in the context of Rastall theory, viz, rotating Rastall black hole that
generalize the Kerr-Newman black hole solution. In turn, we analyze the specific cases of the Kerr-Newman black
holes surrounded by matter like dust and quintessence fields. Interestingly, for a set of parameters and a chosen surrounding
field, there exists a critical rotation parameter ($a=a_{E}$), which corresponds to an extremal black hole with degenerate horizons, while for $a<a_{E}$, it describes a non-extremal black hole with Cauchy and event horizons, and no black hole for $a>a_{E}$ with value $a_E$ is also influenced by these parameters.  We also discuss the thermodynamical quantities associated with rotating Rastall black hole, and analyze the particle motion with the behavior of effective potential. 
\end{abstract}
\maketitle
 
\section{Introduction}
Einstein theory of General Relativity (GR), which is considered as the most beautiful and simplest theory of gravity, admits the covariant conservation of matter energy-momentum tensor. However, since it's formulation people are searching for the alternative theory of gravity and have developed several modified theories of gravity. In this expedition, one of the potential modification to the general theory of relativity was introduced  by P. Rastall \cite{Rastall:1973nw,Rastall:1976uh}, where the usual conservation law of the energy momentum tensor, i.e. $\tensor{T}{^{\mu\nu}_{;\mu}}=0$, is not obeyed. Indeed a non-minimal coupling of matter field to spacetime geometry is considered such that 
\begin{equation}
{T}{{^\nu}_{\mu;\nu}}=\lambda R_{,\mu} \label{Eq1}
\end{equation}
where $\lambda$ quantify the deviation from the Einstein theory of GR, and called the Rastall coupling  parameter. Thus, the divergence of $T^{\mu\nu}$ is proportional to the gradient of the Ricci scalar and that the usual conservation law is recovered in the flat spacetime.  This can be interpreted as an immediate consequence of Mach principle suggesting that the inertia of a local mass depends on the global mass and energy distribution in the universe \cite{Majernik:2006jg}.  Indeed, with this generalization, Einstein could get his famous tensor and thus the corresponding second order equations of motion \cite{3,4}. The Rastall field equation reads \cite{Rastall:1973nw}
\begin{equation}\label{Rastalleq}
G_{\mu\nu}+\kappa\lambda g_{\mu\nu}R=\kappa T_{\mu\nu},
\end{equation}
where $\kappa$ is modified gravitational coupling constant in the Rastall theory. Standard theory of GR can be traced back from Eqs.~(\ref{Eq1}) and (\ref{Rastalleq}) in the limit of $\lambda\rightarrow 0$.  It is one of the most interesting non-conservative theory of
modified gravity  because it provides  an
explanation of the inflation problem, as the simplest
modified gravity scenario to realize the late-time acceleration and other cosmological problems \cite{Campos:2012ez,Fabris:2011wz,Moradpour:2016rml,AlRawaf:1994pn,Batista:2011nu,Carames:2014twa, Salako:2016ihq,Smalley:1974gn, Smalley:1975ry, Wolf:1986wq, Moradpour:2017shy}. Some attention has also been devoted to produce the
static spherically symmetric solutions of the gravitational field equations in the Rastall gravity that includes the neutron star, black hole and worm-holes solutions \cite{Oliveira:2015lka, Moradpour:2016ubd, Moradpour:2016fur,Oliveira:2016ooo,Bronnikov:2016odv}.  In Rastall theory of gravity, the spherically symmetric black hole solution surrounded by perfect fluid was obtained in \cite{Heydarzade:2017wxu}, and  a non-commutative inspired black hole solution in \cite{Ma:2017jko}. The generalization of the black hole solution  sourced by a
Gaussian matter distribution was considered in \cite{Spallucci:2017mto}, while the thermodynamic properties of black hole
solutions in the Rastall gravity were discussed by \cite{Lobo:2017dib}, and demonstrate how the presence of these matter sources may amplify effects caused
by the Rastall parameter in the thermodynamical quantities.  

In this paper, we wish to obtain the rotating counterpart of the spherically symmetric black hole solution obtained in Rastall theory surrounded by a perfect fluid  \cite{Heydarzade:2017wxu}. We begin with a general static, spherically symmetric spacetime,
\begin{equation}
ds^2=-f(r)dt^2+\frac{1}{f(r)}dr^2+r^2d\theta^2+r^2\sin^2\theta d\phi^2,\label{Eq3}
\end{equation} 
where $f(r)$ is a generic metric functional to be obtained by solving Eq.~(\ref{Rastalleq}), for the energy momentum tensor \cite{Heydarzade:2017wxu}
\begin{equation}~\label{EMT}
\tensor{T}{^{\mu}_{\nu}}=\tensor{E}{^{\mu}_{\nu}}+\tensor{\mathcal{ T}}{^{\mu}_{\nu}},
\end{equation}
where $\tensor{E}{^{\mu}_{\nu}}$ is Maxwell's stress tensor of electromagnetic field, given by
\begin{equation}
E^{\mu\nu}=\frac{1}{\mu_0}\left[F^{\mu\alpha}F^{\nu}_{\alpha}-\frac{1}{4}\eta^{\mu\nu}F^{\alpha\beta}F_{\alpha\beta}\right].\label{EMS tensor}
\end{equation}
Assuming the anti-symmetric nature of electromagnetic field tensor $F_{\mu\nu}$, the Maxwell's equations $\tensor{F}{^{\mu\nu}_{;\mu}}=0$ and $F_{[\mu\nu;\gamma]}=0$ leads to 
\begin{equation}\label{EMtensor}
\tensor{F}{^{01}}=\frac{Q}{r^2},
\end{equation} where $Q$ is electrostatic charge, with field tensor in Eq.~(\ref{EMtensor}) the  Maxwell's tensor (\ref{EMS tensor}) leads to the following traceless tensorial form
\begin{equation}
\tensor{E}{^{\mu}_{\nu}}=\frac{Q^2}{kr^4}[-1,-1,1,1].
\end{equation}
The energy momentum tensor must obey the symmtery property of modified Rastall tensor in Eq. (\ref{Rastalleq}) and Schwarzschild like symmetry in (\ref{Eq3}), therefore the stress tensor of surrounding perfect fluid reads \cite{Ghosh:2015ovj, Kiselev:2002dx}
\bea
& \mathcal{T}^0_0=\mathcal{T}^1_1=-\rho_s,\nn\\
& \mathcal{T}^2_2=\mathcal{T}^3_3=\frac{1}{2}(1+3\omega_s)\rho_s.
\eea
Solving the modified Einstein's equation (\ref{Rastalleq}) with energy-momentum tensor Eq.~(\ref{EMT}), we obtain \cite{Heydarzade:2017wxu}
\begin{equation}
f(r)=1-\frac{2M}{r}+\frac{Q^2}{r^2}-\frac{N_s}{r^{\xi}},\label{lapse}
\end{equation}
with 
\beq
\xi={\frac{1+3\omega_s-6\kappa\lambda(1+\omega_s)}{1-3\kappa\lambda(1+\omega_s)}}.\label{xieq}
\eeq
Here, $M$ and $N_s$ are integration constant representing the black hole mass and structure parameter for surrounding field, respectively.
The energy density of surrounding fluid has a form \cite{Heydarzade:2017wxu}
\begin{equation}\label{density}
\rho_s(r)=-\frac{3\mathcal{W}_sN_s}{\kappa r^{(2\xi-3\omega_s+1)}},
\end{equation}
where $\mathcal{W}_s$ is a geometric constant define as,
\begin{equation}
\mathcal{W}_s=-\frac{(1-4\kappa\lambda)\big(\kappa\lambda(1+\omega_s)-\omega_s\big)}{(1-3\kappa\lambda(1+\omega_s))^2}.
\end{equation}
$\mathcal{W}_s$ depends on the Rastall geometric parameter $\kappa\lambda$ as well as on the field state parameter $\omega_s$. 
The aim of this paper is to obtain and study the rotating counter part of the static solution Eq.~(\ref{Eq3}), i.e. to obatin the Kerr-Newman (KN) metric like solution of Rastall theory. The KN black hole is one of the most extraordinary solution of Einstein-Maxwell equation that represents the prototype charged black hole that can result from the gravitational collapse \cite{Nathanail:2017wly}. The rotating black hole enjoy many interesting properties differ from it's non-rotating counterpart \cite{Atamurotov:2015xfa, Ghosh:2014pba, Ghosh:2013bra}. It turns out that spin plays important role in astrophysical process, and non-rotating black hole can not be tested by astrophysical observations \cite{Bambi:2011mj}. This motivated us to seek generalization of Rastall black hole obtained in \cite{Heydarzade:2017wxu} to axially symmetric case or to KN-like solution surrounded by perfect fluid in Rastall theory. The obtained rotating black hole solution revert back to Kerr and KN solution in appropriate limits.
\section{Kerr-Newman solution in Rastall theory}
In order to obtain a four parameters regulated axially symmetric black hole metric in Rastall theory that has mass ($M$), electric charge ($Q$), spin parameter ($a$), and the Rastall parameter ($\lambda$); which is a rotating counterpart of static, spherically symmetric black hole solution of the Rastall theory of gravity, we employ the standard Newman\(-\)Janis algorithm \cite{nja} on solution (\ref{Eq3}). Firstly, we perform the Eddington\(-\)Finkelstein Coordinate (EFC) transformation $du = dt - f(r)^{-1} dr,$ such that the metric can be written down in terms of advance null coordinates, which takes the following form
\begin{eqnarray}
ds^2 =  -f(r) du^2 - 2 dudr + {r^2}d\Omega^2. \label{SchwEF1}
\end{eqnarray}
Furthermore, we can write the inverse metric in terms of complex null tetrad $ Z^\mu = (l^{\mu},\;n^{\mu},\;m^{\mu},\;\bar{m}^{\mu}$), as
\begin{equation}
{g}^{\mu\nu} = l^{\mu} n^{\nu} +  l^{\nu} n^{\mu} - m^{\mu} \bar{m}^{\nu}-\bar{m}^{\mu} {m}^{\nu},
\label{NPmetric}
\end{equation}
where null tetrad are
\begin{eqnarray*}
l^{\mu} &=& \delta^{\mu}_r,\\
n^{\mu} &=&  \delta^{\mu}_u - \frac{1}{2} f(r)\delta^{\mu}_r ,\\
 m^{\mu} &=& \frac{1}{\sqrt{2}r}   \left( \delta^{\mu}_{\theta}
  + \frac{i}{\sin\theta} \delta^{\mu}_{\phi} \right).
\end{eqnarray*}
The null tetrad are orthonormal and obey the following defining conditions, namely that all the vectors of the tetrad have zero magnitude
\begin{eqnarray}
l_{a}l^{a} = n_{a}n^{a} = ({m})_{a} ({m})^{a} = (\bar{m})_{a} (\bar{m})^{a}= 0,  \nonumber \\
l_{a}({m})^{a} = l_{a}(\bar{m})^{a} = n_{a}({m})^{a} = n_{a}(\bar{m})^{a}= 0, \; \nonumber \\
l_a n^a = 1, \; ({m})_{a} (\bar{m})^{a} = 1.
\end{eqnarray}
 Following the Newman\(-\)Janis prescription
\cite{nja}, we allow coordinates to have complex values, whereas in order to get real $l^{\mu}$ and $n^{\mu}$ we can choose the following  transformation
\begin{equation}\label{transf}
{x'}^{\mu} = x^{\mu} + ia (\delta_r^{\mu} - \delta_u^{\mu})
\cos\theta \rightarrow \\ \left\{\begin{array}{ll}
u' = u - ia\cos\theta, \\
r' = r + ia\cos\theta, \\
\theta' = \theta, \\
\phi' = \phi. \end{array}\right.
\end{equation}
where $"a"$ is a newly introduced parameter, which will be further related to spin of black hole. The null tetrad transforms as a vector and undergoes a
transformation $Z'^a = Z^b{\partial x'^a}/{\partial x^b} $, thus we obtain
\begin{eqnarray}\label{NPkerr}
l^{\mu} &=& \delta^{\mu}_r, \nonumber \\
n^{\mu} &=& \left[ \delta^{\mu}_u - \frac{1}{2} \tilde{f}(r,\theta) \delta^{\mu}_r \right], \nonumber \\
 m^{\mu} &=& \frac{1}{\sqrt{2}(r+ia\cos\theta)} \times  \left(ia(\delta^{\mu}_u-\delta^{\mu}_r)\sin\theta + \delta^{\mu}_{\theta} + \frac{i}{\sin\theta} \delta^{\mu}_{\phi} \right),
 \end{eqnarray}
where $\tilde{f}(r,\theta)$ is the complexified form of function $f(r)$.  Using tetrad Eq.(\ref{NPkerr}) and following Eq.(\ref{NPmetric}), the non-zero component of the inverse of new metric can be written as
\begin{eqnarray}\label{KCinvm}
&& g^{u u } = \frac{a^2 \sin^2(\theta )}{ \Sigma (r ,\theta )}, \quad 
g^{u \phi} = \frac{a}{  \Sigma  (r ,\theta )} , \nonumber  \\ \nn
&& g^{u r } =-1-\frac{a^2 \sin ^2(\theta )}{  \Sigma  (r, \theta)}\\ \nn
&& g^{rr} = \frac{a^2 \sin ^2 \theta }{  \Sigma  (r, \theta)} + \tilde{f}(r,\theta),\quad g^{r \phi} = -\frac{a}{ \Sigma  (r, \theta)}\\ \nn
&&g^{\theta \theta} = \frac{1}{  \Sigma  (r,\theta )},\quad g^{\phi \phi} = \frac{1}{   \Sigma  (r, \theta ) \sin^2 \theta } \nn
\end{eqnarray}
where
\begin{equation}
\tilde{f}(r,\theta) = 1 - \frac{2Mr}{\Sigma(r,\theta)}+\frac{Q^2}{\Sigma(r,\theta)}-\frac{N_s}{\Sigma(r,\theta)^{\xi/2}}.
\end{equation}
From the transformed tetrad the new line element takes the following form
\begin{align}
ds^2  &= -\tilde{f}(r,\theta)du^2 - 2dudr -2\left(1-\tilde{f}(r,\theta)\right)a\sin^2\theta du d\phi + 2a\sin^2\theta dr d\phi +\Sigma(r,\theta) d\theta^2\nn\\
 &  +\sin^2\theta\left[\Sigma(r,\theta)+(2-\tilde{f}(r,\theta))a^2\sin^2\theta\right] d\phi^2 .
 \label{metric002}
\end{align}
To write the metric in the Boyer-Lindquist Coordinates (BLC) form, we make further coordinate transformation  
\begin{eqnarray}\label{transfer}
&&du=dt+\psi(r)dr,\quad d\phi=d\phi+\chi(r)dr,
\end{eqnarray}
where the two functions $\psi(r)$ and $\chi(r)$ are chosen as to eliminate the $g_{tr}$
and $g_{r\phi}$ components. Thus plug Eq. (\ref{transfer}) into Eq. (\ref{metric002}) and demanding all off-diagonal component to be zero except $g_{t\phi}$ (to preserve axial symmetry), its turn out that $\psi(r)$ and $\chi(r)$ are now functions of both $r$ and $\theta$, and have following form
\begin{eqnarray}
&&\psi(r,\theta)=\frac{\Sigma(r,\theta)+a\sin^2\theta}{\left(\Sigma(r,\theta)\tilde{f}(r,\theta)+a\sin^2\theta\right)},\nonumber\\
&&\chi(r,\theta)=\frac{-a}{\left(\Sigma(r,\theta)\tilde{f}(r,\theta)+a\sin^2\theta\right)}.
\end{eqnarray}  
This $\theta$ dependence in EFC to BLC transformation can be attributed to the fact that we are dealing with non-vacuum surrounding and a modified theory of gravity \cite{Bambi:2013ufa, Azreg-Ainou:2014pra}. 
 Henceforth, we will omit writing the dependency on $\theta$ and $r$ in the functions $\Sigma$ as well as in $\Delta$ (which is defined further). Having these two expressions for  $\psi(r,\theta)$ and $\chi(r,\theta)$, we finally obtain the rotating KN-like black hole in the BLC form in the context of  Rastall theory, which reads as
\begin{align}
ds^2 &=-\left(\frac{\Delta-a^2\sin^2\theta}{\Sigma}\right){dt}^2 -2\left(1-\frac{\Delta-a^2\sin^2\theta}{\Sigma}\right)a\sin^2\theta {dt}d\phi +\frac{\Sigma}{\Delta}{dr}^2\nn \\
  & +\Sigma d\theta^2+\sin^2\theta\left[\Sigma+\left(2-\frac{\Delta-a^2\sin^2\theta}{\Sigma}\right)a^2\sin^2\theta\right]d\phi^2,
\label{metric2}
\end{align}
with
$\Sigma=r^2+a^2\cos^2\theta $, and $\Delta= r^2+a^2-2 M r+Q^2-\frac{N_s}{\Sigma^{(\xi-2)/2}}$.  Note that the metric Eq.~(\ref{metric2}) matches with the various known black hole solutions in the suitable limits. It include the KN spacetime \cite{Newman:1965my} as the special case when the black hole is surrounded by the vacuum ($N_s\rightarrow 0$), and perfectly reduces to Kerr solution \cite{Kerr:1963ud} in the limit ($N_s, Q\rightarrow 0$). Even in the GR limit of rotating Rastall black hole spacetime, i.e., vanishing Rastall coupling ($\lambda\rightarrow 0$), Eq.~(\ref{metric2}) mimic the KN black hole solution in the presence of quintessence ($\xi=3\omega_s+1$), while the well known rotating quintessential black hole in GR \cite{Ghosh:2015ovj} can be obtained by further choosing $Q=0$. Reissner-Nordstrom (RN) solution can be obtained on further restricting both ($N_s, a\rightarrow0$).\newline
It is well known in GR that the spherically symmetric static black hole and its rotating counterpart have the same source, if it exit, i.e., vacuum for both Schwarzschild and Kerr black hole while charge for RN and KN black holes. But in modified theories or black holes with non-trivial source an additional stress generates corresponds to rotating fluid. Hence, the source for the static and rotating black hole solution are different with rotating solution has additional stresses \cite{Ghosh:2014hea, Carmeli:1975kg}. Further, the Newman\(-\)Janis algorithm is well defined for GR, it can be applied to the Rastall gravity as well. This is because the Rastall gravity is identical to the GR through a redefined energy momentum tensor \cite{Visser:2017gpz}. In order to calculate the explicit form of the energy-momentum tensor compatible with the modified Einstein equation (\ref{Rastalleq}) and for mathematical simplicity we consider a locally non-rotating observer with following orthonormal tetrad of 4-vectors \cite{Bardeen:1972fi} 
\begin{equation}
e^{(a)}_{\mu}=
\begin{pmatrix}
 \sqrt{-(g_{tt}-(g_{t\phi})^2/g_{\phi\phi})} & 0& 0 &0\\
0 & \sqrt{ g_{rr}}& 0 &0\\
0 & 0& \sqrt{ g_{\theta\theta}} &0\\
g_{t\phi}/\sqrt{g_{\phi\phi}} & 0& 0 &\sqrt{ g_{\phi\phi}}\\
\end{pmatrix}.\nn
\end{equation}
The components of the energy momentum tensor in the orthonormal frame read $T^{(a)(b)}=e^{(a)}_{\mu}e^{(b)}_{\nu}(G^{\mu\nu}+\kappa\lambda Rg^{\mu\nu})$. The expressions for the components of the energy momentum tensor  for rotating Rastall black hole are clumsy although finite. In particular, at equatorial plane ($\theta=\pi/2$) they took considerably simple form as follow 

\begin{eqnarray}
T^{(0)(0)}&=&-\frac{2\kappa\lambda}{r^2}\left(rm''+2m'\right)+\frac{2(r^2+a^2)^2m'-a^2r\Delta m''}{r^3(r^3+a^2r+2a^2m)},\nn\\
T^{(1)(1)}&=&\frac{2r\kappa\lambda m''+2(\kappa\lambda-1)m'}{r^2},\nn\\
T^{(2)(2)}&=&\frac{r(2\kappa\lambda-1)m''+4\kappa\lambda m'}{r^2},\nn\\
T^{(3)(3)}&=&\frac{2\kappa\lambda}{r^2}\left(rm''+2m'\right)+\frac{2a^2\Delta m'-r(r^2+a^2)^2m''}{r^3(r^3+a^2r+2a^2m)},\nn\\
T^{(0)(3)}&=&\frac{-a(r^2+a^2)(rm''-2m')\sqrt{\Delta}}{r^3(r^3+a^2r+2a^2m)},
\end{eqnarray}
with $m=\left.\left( M-Q^2/2r+N_s/{2r\Sigma^{(\xi-2)/2}}\right)\right|_{\theta=\pi/2}$ and $'$ represents derivative with respect to $r$. We have demonstrated that the rotating black hole solution in the Rastall gravity is supported by the energy momentum tensor, the components of which die off very rapidly at large $r$. This happens to all rotating black hole solution where source is non-trivial, e.g. \cite{Ghosh:2014hea, Carmeli:1975kg}. The non-vanishing off-diagonal component $T^{(0)(3)}$ represents the matter tangential flow at equatorial plane, which is as expected clearly zero for non-rotating case. In the GR limit ($\lambda=0$) for black hole surrounded by vacuum ($N_s=0$), the components of energy momentum tensor at pole along symmetric axis take the following form 
\begin{eqnarray}
T^{(0)(0)}&=& \frac{2r^2m'}{(r^2+a^2)^2}=-T^{(1)(1)},\nn\\
T^{(2)(2)}&=&-\frac{r(r^2+a^2)m''+2a^2 m'}{(r^2+a^2)^2}=T^{(3)(3)},\nn
\end{eqnarray}
 which are in full consistent with the results shown in  \cite{Neves:2014aba}; $T^{(a)(b)}$ are diagonal in this case.

Likewise standard KN metric in the BLC form, the modified metric Eq.~(\ref{metric2}) is also independent of $\phi$ and $t$. Therefore, the metric admits two Killing vectors associated with the time translation invariance and rotational invariance defined as $\eta_{(t)}^{\mu}=\delta^{\mu}_t$ and $\eta_{(\phi)}^{\mu}=\delta^{\mu}_{\phi}$ respectively. Thus, by definition of Killing vectors, momentum associated with translation along $t$ and $\phi$ coordinates are constant of motion. Unlike the Kerr spacetime, the spacetime geometry of metric (\ref{metric2}) is not Ricci flat and henceforth the black hole (\ref{metric2}) will be called rotating Rastall black hole. We start in the next section with the  horizons structure  of the rotating Rastall black hole,  which make them distinguishable from their GR counterparts.
\section{Properties of horizons}
In this section, we discuss the physical properties of rotating Rastall black hole obtained in the previous section. Likewise the KN metric, the rotating Rastall black hole metric (\ref{metric2}) is also singular at $ \Sigma = 0 $ and at $\Delta=0$. The solution of $\Sigma=0$ is a ring shape physical singularity at the equatorial plane of the center of rotating black hole with radius $a$. On the other hand, $\Delta=0$ is a coordinate singularity which determines the horizon associated with spacetime, horizon in terms of (${t, r, \theta, \phi}$) is a null hypersurface of constant $r$, i.e. 
\begin{equation}\label{horizon}
 g^{\mu\nu}\partial _{\mu}r\partial_{\nu}r=0\;\;\; or\;\; g^{rr}=\Delta=0, 
\end{equation} 
 where $\partial _{\mu}r$ is the normal to the said hypersurface. Thus, horizons are zeros of $\Delta=0$ 
\beq
r^2+a^2-2 M r+Q^2-\frac{N_s}{\Sigma^{(\xi-2)/2}}=0,\label{horizon1}
\eeq
which are $\theta$ dependent and different from the KN black hole horizon, where it is $\theta$ independent. This $\theta$ dependency is due to the non-minimal coupling of surrounding field with geometry.  The Eq.~(\ref{horizon1}) admits some real solutions which corresponds to the radial coordinate of Cauchy horizon ($r_-$), event horizon ($r_+$) and cosmological horizon ($r_q$ if present) with $r_-\leq r_+ \leq r_q$, whereas each horizon radius depend on $a, Q, N_s, \kappa\lambda, \theta$ and $M$.\\
It must be noted from Eq. (\ref{xieq}) that ($\xi=2$) is not allowed for any value of Rastall parameter except for surrounding radiation field ($\omega_s=1/3$). 

A Killing horizon of a rotating black hole is the null hypersurface where the linear combination of Killing vector for time translation and rotation symmetry is null, $\chi^{\mu} \chi_{\mu} =0$ with $\chi^{\mu}=\eta_{(t)}^{\mu}+\Omega \eta_{(\phi)}^{\mu}$.  The outer edges of ergosphere corresponds to the timelike hypersurface where the time translational Killing vector become null, i.e. $g_{tt}=0$ .  A priori, ergosphere and event horizon are distinct hyper-surfaces \cite{Ghosh:2015ovj} and for the standard KN black hole these surfaces reads
\begin{eqnarray}
r^{+} & = & M + \sqrt{M^2-a^2-Q^2},  \nonumber \\
r^{ergo} & = & M + \sqrt{M^2-a^2 \cos^2\theta -Q^2}. 
\end{eqnarray}
For rotating Rastall black hole the radial coordinates of static limit surface are the zeros of 
\begin{equation}
\eta_{(t)}^{\mu} \eta_{(t)\mu}^{}=g_{tt}=-\left(\frac{\Delta-a^2\sin^2\theta}{\Sigma}\right)
\end{equation} 
Next, in the context of Rastall gravity, we are examining the two special cases of surrounding fields, namely the dust and quintessence. 
\subsection{The rotating Rastall black hole surrounded by dust field}
In the case of dust, the state parameter is $\omega_s=0$ and the $\xi$ reduces to
\begin{equation}
\xi=\frac{1-6\kappa\lambda}{1-3\kappa\lambda}.\label{xi-dust}
\end{equation}
Next, we compare this rotating Rastall black hole surrounded by dust with the KN black hole surrounded by quintessence \cite{Xu:2016jod}. In the limiting case $\lambda\rightarrow 0$ one obtain 
\beq
\Delta=r^2+a^2-2Mr+Q^2-N_s\Sigma^{1/2};
\eeq
Unlike, the non-rotating case where one obtain the RN solution in the limit $\lambda\rightarrow 0$, we do not get the KN black hole due to presence of extra term $N_s$. Demanding the positivity of surrounding field energy density (weak energy condition), from Eq.~(\ref{density}) we must have $\mathcal{W}_sN_s\leq 0.$ Thus, positive field structure constant ($N_s>0$) restrict the value of deviation parameter by $0\leq\kappa\lambda\leq 1/4$, otherwise $\kappa\lambda$ will be out of this range for negative structure constant. On the other hand, to obtain the rotating Kiselev metric \cite{Ghosh:2015ovj,Kiselev:2002dx}, the effective state parameter for the surrounding dust field modified in Rastall theory reads as
\begin{equation}
\omega_{eff}=\frac{1}{3}\left(-1+\frac{1-6\kappa\lambda}{1-3\kappa\lambda}\right).\label{omegaeff}
\end{equation}
\begin{figure*}
 \begin{tabular}{c c}
\includegraphics[scale=0.7]{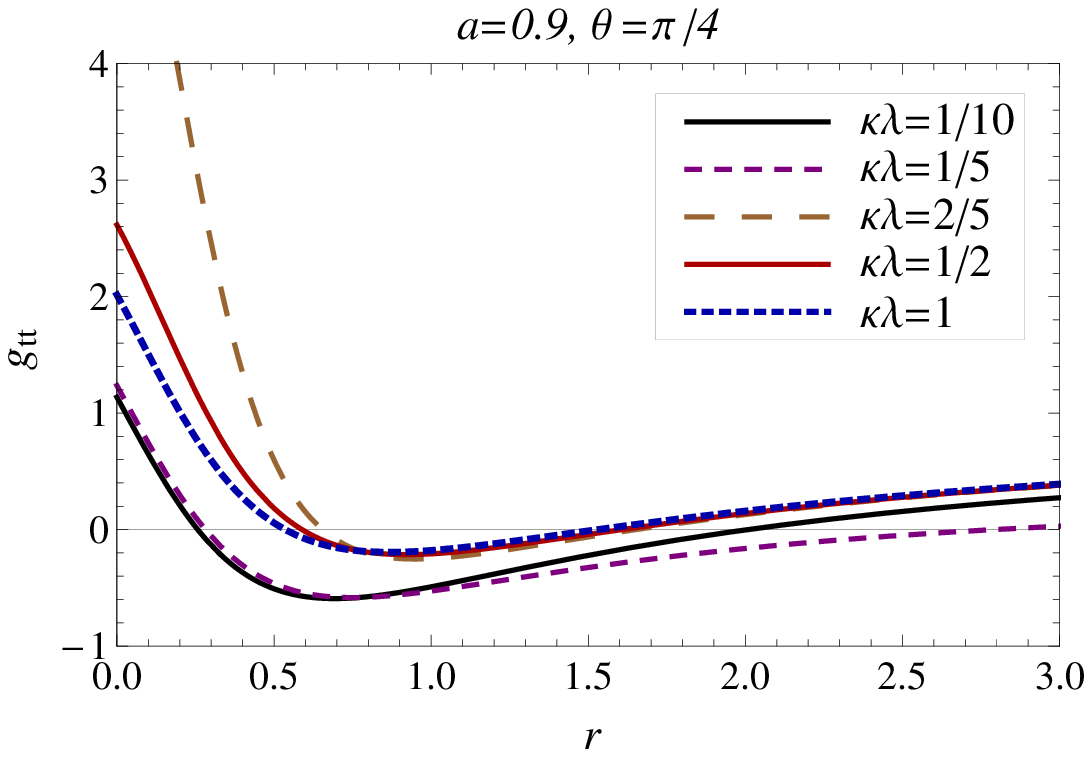}&
\includegraphics[scale=0.7]{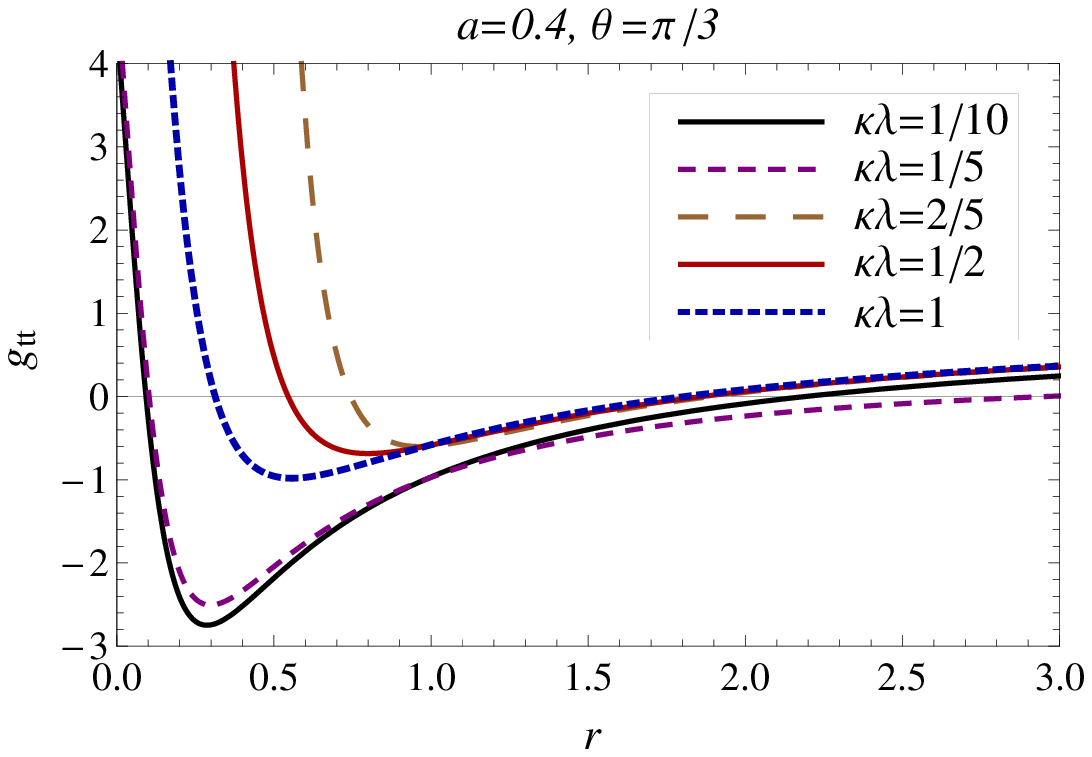} 
\end{tabular}
\caption{Plot showing the variation of $g_{tt}$ with $r$ for different values of parameter $\kappa\lambda$ for black hole ($M=1, Q=0.2$) surrounded by dust field.}
\label{SLSdust}
\end{figure*}
\begin{figure*}
 \begin{tabular}{c c}
\includegraphics[scale=0.7]{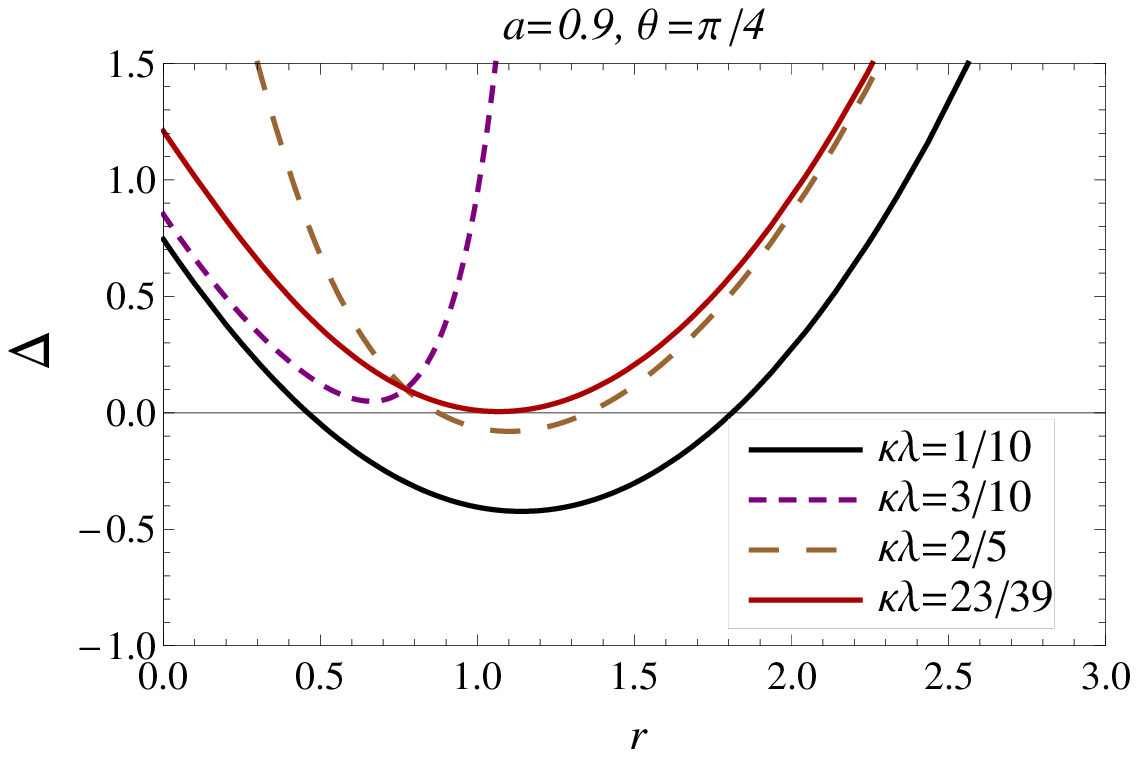}&
\includegraphics[scale=0.7]{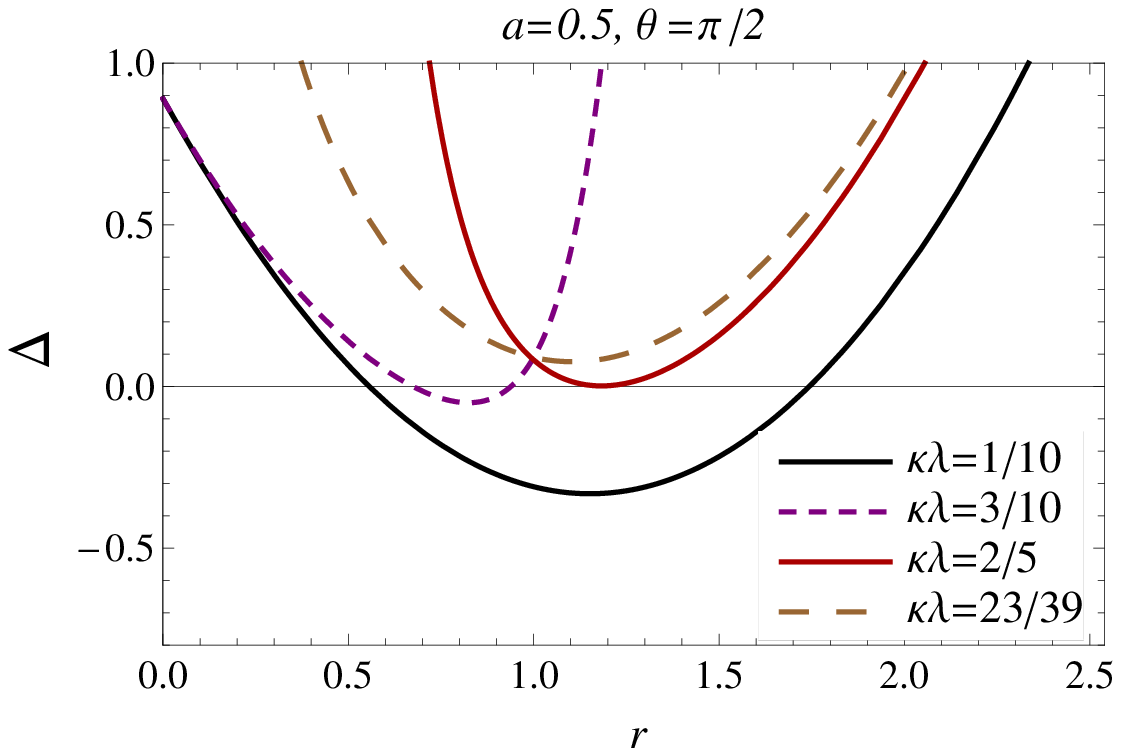} \\
\includegraphics[scale=0.7]{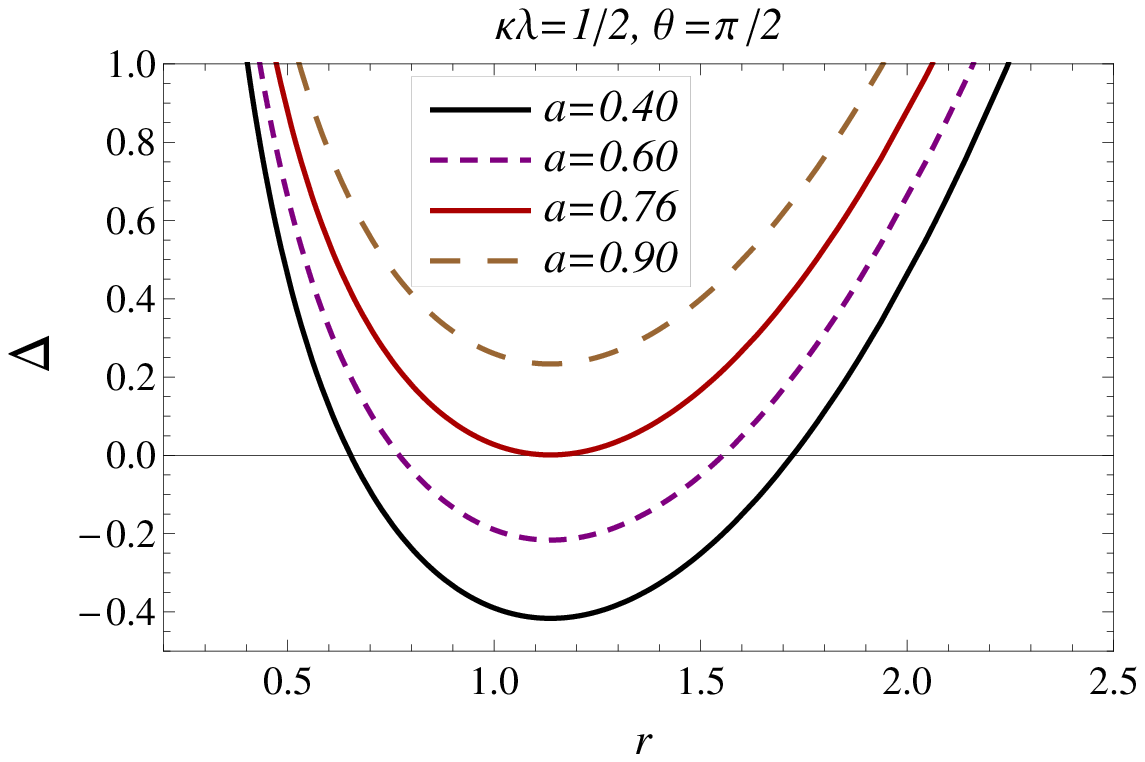}&
\includegraphics[scale=0.7]{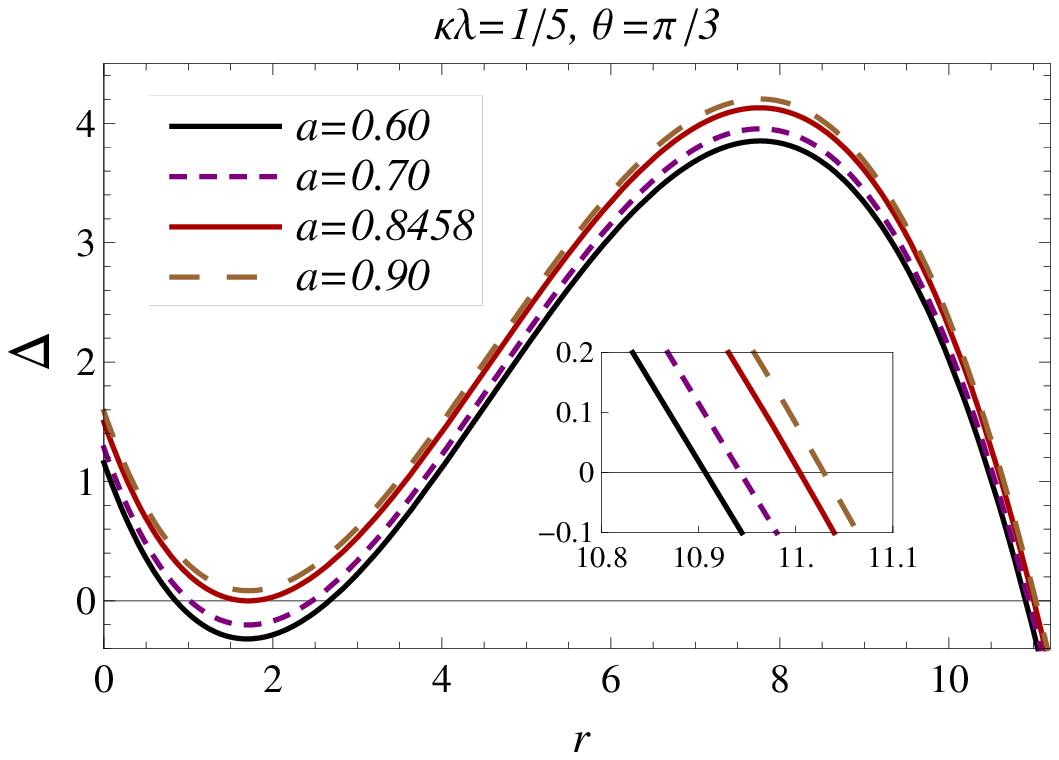}
\end{tabular}
\caption{Plot showing the variation of $\Delta$ with $r$ for different values of parameter $\kappa\lambda$ and $a$ for black hole ($M=1, Q=0.2$) surrounded by dust field. (\textit{Red solid curve}) corresponds to the extremal black hole.}
\label{Horizondust}
\end{figure*}
This effective state parameter is always non-zero unless we reach in GR limit $\kappa\lambda=0$. Starting with the surrounding dust field $(\omega_s=0)$ in GR we end up with black hole solution surrounded by effectively different field of state parameter $\omega_{eff}$ in Rastall theory. Surprisingly, depending upon the numerical value of deviation parameter $\kappa\lambda$, the effective parameter dictates various surrounding field and this is extensively discuss in \cite{Heydarzade:2017wxu}. The behavior of  ergosphere is shown in Fig.~\ref{SLSdust}.  
The horizon radius of a KN black hole in vacuum is less than the radius of KN black hole surrounded by dust field $$r_{KN}=\frac{(2M+N_s)\pm \sqrt{(2M+N_s)^2-4(a^2+Q^2)}}{2},$$ even in the limit $\lambda\rightarrow 0$.  Clearly, the surrounding dust field increase the horizon radius of isolated black hole subject to the choice of positive structure parameter, e.g.,  for particular values of $M=1$, $a=0.8$, and  $Q=0.2$, horizon radii of  KN black hole  are ($	r_-=0.434315,\quad$ $r_+=1.56569$) whereas for the KN black hole surrounded by dust field with  $N_s=2$, the horizon radii become  ($	r_-=0.177913,\quad$ $r_+=3.82209$).  Furthermore, if we consider a non-minimal coupling between geometry and surrounding field then it is difficult to get the analytic solutions of Eq.~(\ref{horizon1}). However, from numerical analysis we can check that for fixed values of $M$, $a$ and $Q$, the horizon radius depends upon the field structure parameter $N_s$ and modified gravitational coupling $\kappa\lambda$.  In Fig. \ref{Horizondust}, we have plotted the $\Delta$ with radial coordinate $r$ for a set of parameter, for both with and without cosmological horizon. In this case, the numerical analysis of $\Delta=0$ implies the existence of upto three roots which correspond to the horizons of a rotating Rastall black hole spacetime  (\ref{metric2}). An extremal black hole occurs when $\Delta=0$ has a double root, corresponds to the degenerate horizon, i.e., two horizons coincide. When $\Delta=0$ has no root, i.e., no horizon exists, which mean there is no black hole (cf. Fig.~\ref{Horizondust} ). We have explicitly shown that, for each $\kappa\lambda$,  one gets two horizons for $a< a_E$ (subject to suitable $\kappa\lambda$), and when $a=a_{E}$ the two horizons coincide, i.e., we have an extremal black hole with degenerate horizons (Fig.~\ref{Horizondust}).
\begin{table}
 \centering
\begin{tabular}{ c  c  c  c  c  c}
 \hline
  $\kappa\lambda$     &     $\xi$     & $N_s$ &  $r_-$ &  $r_+$ & $r_q$ \\
 \hline
    $\frac{1}{10}$    &      $\frac{4}{7}$     & +2 & 0.239092  & 6.29064 & -\\
    \hline
    $\frac{1}{8}$ & $\frac{2}{5}$ & +2 & 0.258577  & 9.79379 & -\\
  \hline
  $ \frac{3}{10}$ & -8 & -2 & 0.4347412  & 0.824133 & -\\
  \hline
   $\frac{1}{5}$ &    $-\frac{1}{2}$ & +0.2 & 0.4150  & 2.5445 & 20.4153\\
  \hline
   $\frac{2}{9}$ &    $\frac{7}{3}$ & +0.1 & 0.4274  & 2.140 & 7.43207\\
  \hline
   $\frac{1}{4}$ &    -2 & +0.05 & -  & - & 0.432767\\
   \hline
  \end{tabular}
    \caption{Table showing horizon radii of a charged rotating black hole ($M=1, a=0.8, Q=0.2$) surrounded by dust in Rastall theory for various gravitational coupling strength $\kappa\lambda$. The outermost cosmological horizon ($r_q$) is present for those values of $\kappa\lambda$ which leads to the SEC violation.}
  \label{dusttable}
\end{table}
Indeed we have degenrate horizons for extremal value of $a(=a_E)$ or $\kappa\lambda$.
From Eqs.~(\ref{horizon1}),(\ref{xi-dust}), we calculate the horizon radii of charged-rotating black hole in Rastall theory and comparison with the corresponding radius in GR is shown in table \ref{dusttable}. This is evident from Eq. (\ref{omegaeff}) that for a suitable choice of $\kappa\lambda$, the effective state parameter may leads to the violation of SEC and a de-Sitter type outer cosmological horizon ($r_q$) will be present e.g., $\kappa\lambda=1/5$ leads to the $\omega_{eff}<-1/3$ and a cosmological horizon is present even in the case of surrounding dust field (cf. Fig.  \ref{Horizondust}).
 \subsection{The rotating Rastall black hole surrounded by quintessence field}
\begin{figure*}
 \begin{tabular}{c c}
\includegraphics[scale=0.7]{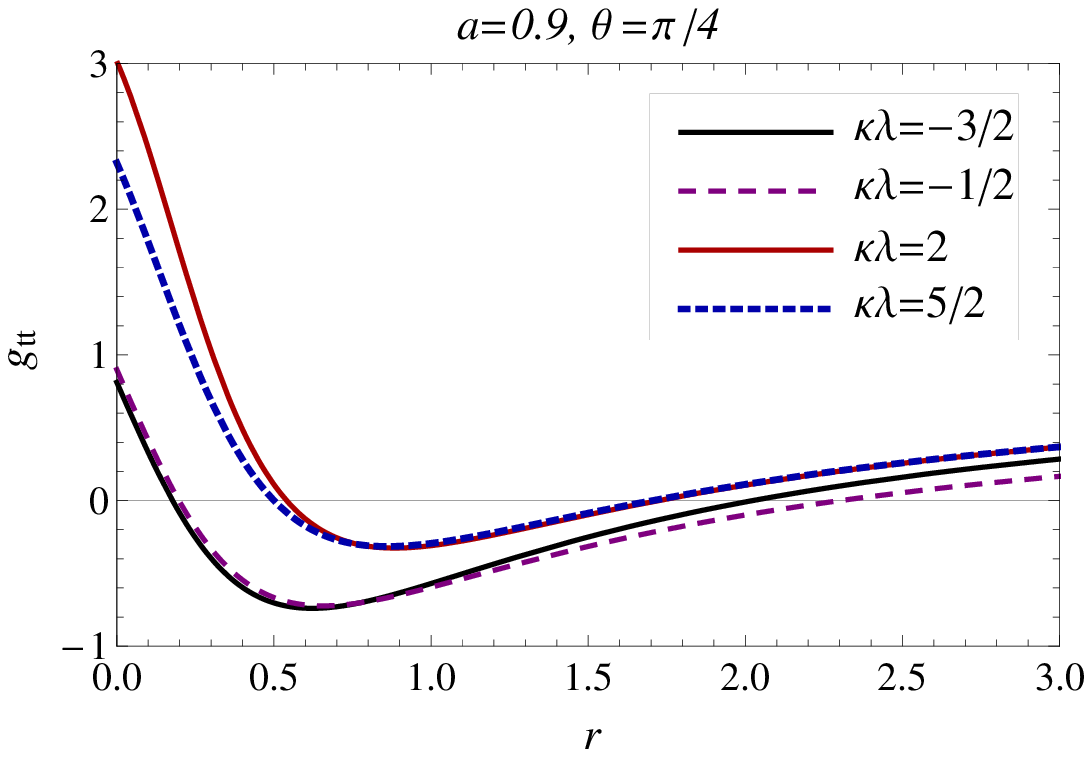}&
\includegraphics[scale=0.7]{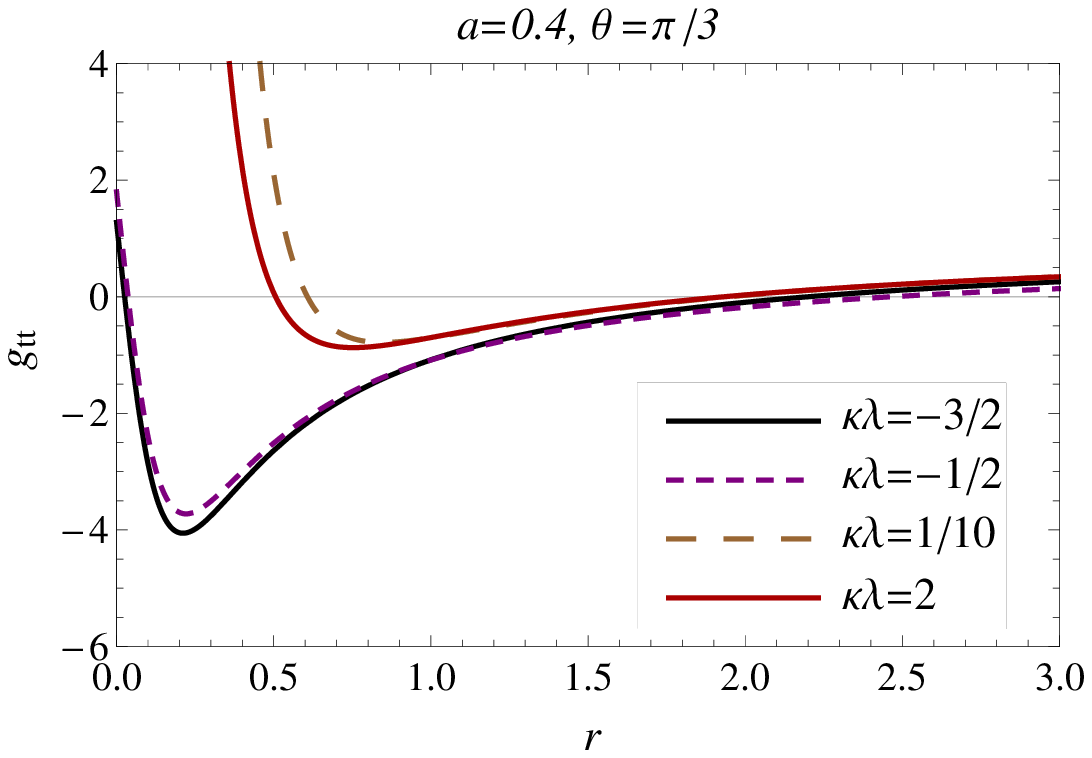} 
\end{tabular}
\caption{Plot showing the variation of $g_{tt}$ with $r$ for different values of parameter $\kappa\lambda$ for black hole surrounded by quintessence field. }
\label{SLSquint}
\end{figure*}

 The rotating Kiselev black hole solution has been analyzed by Ghosh \cite{Ghosh:2015ovj} which is a  Kerr black hole  surrounded by quintessence and later it was extended for charged case in \cite{Xu:2016jod}. The rotating Rastall black hole surrounded by quintessence field is characterized by equation of state parameter $(-1<\omega_s<-1/3)$, in particular we choose $\omega_s=-2/3$~\cite{Kiselev:2002dx, Xu:2016jod}. In the presence of such field, the spacetime is no longer asymptotically flat and inflating Universe has a cosmological horizon due to negative pressure of quintessential field. The field parameter and effective sate parameter for quintessence surrounded black hole in Rastall theory take the form
\begin{equation}
\xi=\frac{-1-2\kappa\lambda}{1-\kappa\lambda}, \,\,\,\,\,\ \omega_{eff}=\frac{1}{3}\left(-1-\frac{1+2\kappa\lambda}{1-\kappa\lambda}\right).
\end{equation}
By comparing the corresponding solution in GR \cite{Ghosh:2015ovj}, we can easily see the distinction between these two solutions. In the zero Rastall coupling limit ($\lambda\rightarrow 0$), we can traced back the GR solution. The geometric parameter $\mathcal{W}_s$ takes the form
\begin{equation}
\mathcal{W}_s=-\frac{(1-4\kappa\lambda)(2+\kappa\lambda)}{3(1-\kappa\lambda)^2}.
\end{equation}
Condition to the validation of weak energy condition, $N_s$ takes different values corresponding to the choice of different values of $\kappa\lambda$. If we compute effective equation of state parameter then we can realize that it will never be $-2/3$ unless $\kappa\lambda=0.$ For $-1/2\leq \kappa\lambda<1$, surrounding field (with $\omega_{eff}\leq-1/3$) violating SEC, will contribute to the accelerating expansion of the universe but with different strength as compare to quintessence field.  On contrary, for $\kappa\lambda\leq-1/2 $ or $\kappa\lambda>1$ the surrounding quintessential field in Rastall theory will respect the SEC, and will regulate the decelerating expansion of Universe and eventually play a role in it's contraction. Behavior of ergosurface for different Rastall coupling strength is shown in Fig.~\ref{SLSquint}.
\begin{figure*}
 \begin{tabular}{c c}
\includegraphics[scale=0.7]{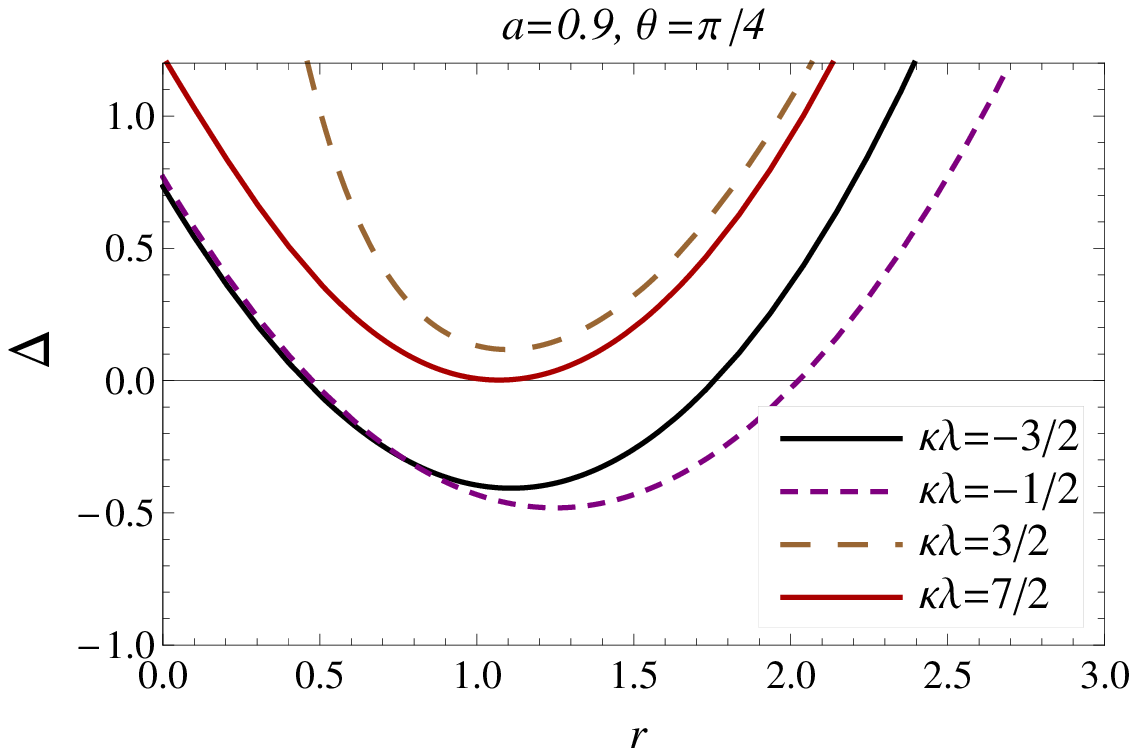}&
\includegraphics[scale=0.7]{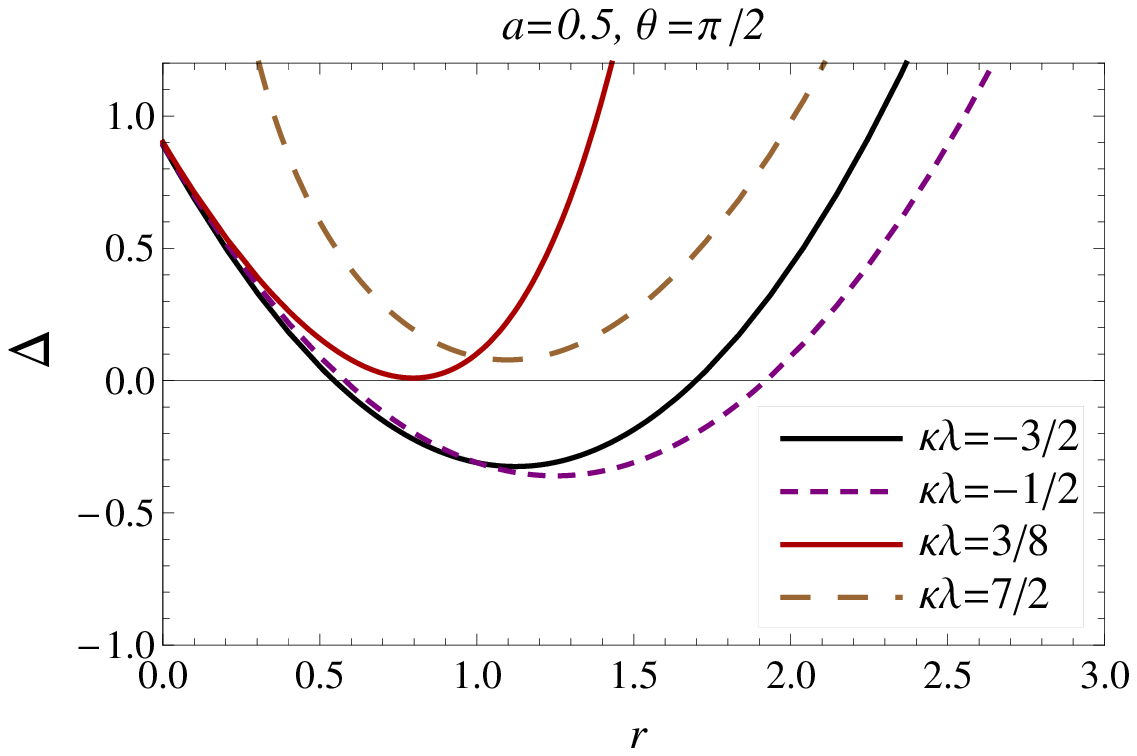}\\
\includegraphics[scale=0.7]{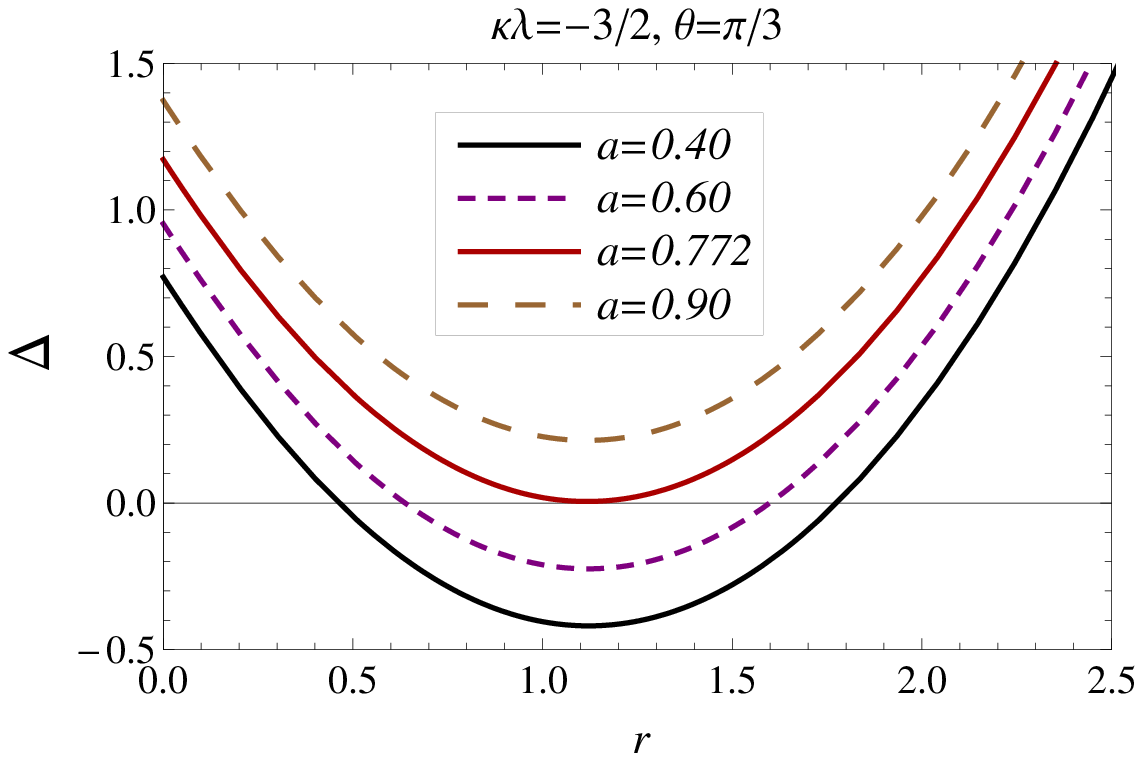}&
\includegraphics[scale=0.7]{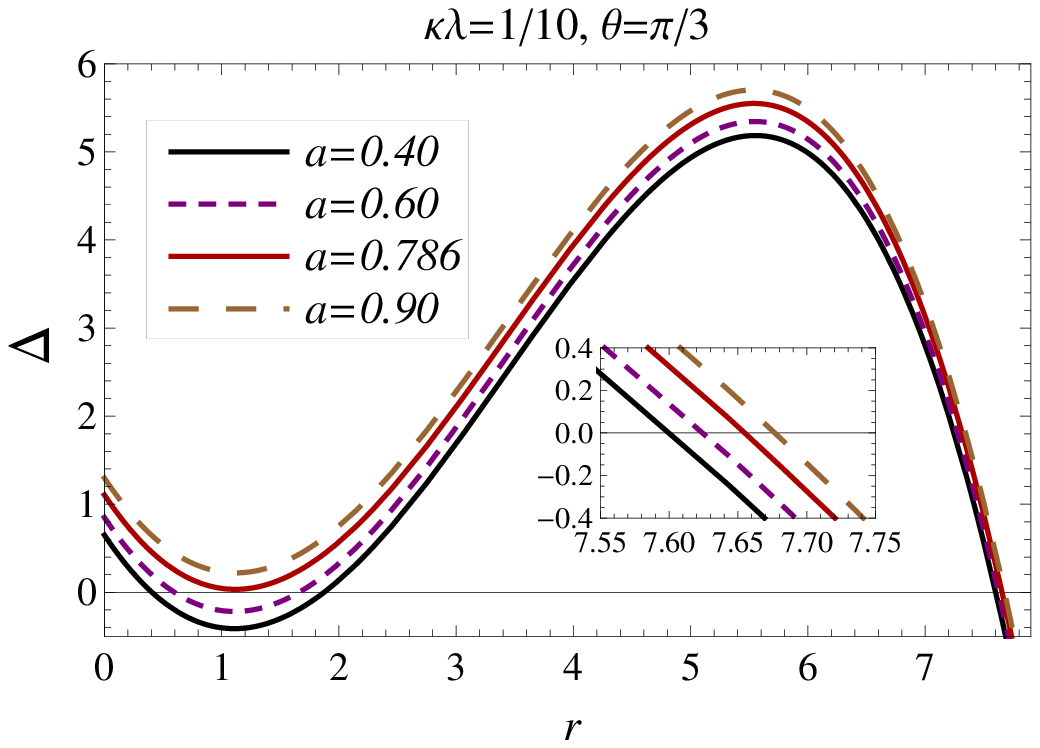}
\end{tabular}
\caption{Plot showing the variation of $\Delta$ with $r$ for different values of parameter $\kappa\lambda$ and $a$ for rotating black hole surrounding by quintessence in Rastall theory. (\textit{Red solid curve}) corresponds to the extremal black hole.}
\label{Horizonquint}
\end{figure*}

\begin{figure}
\includegraphics[scale=0.7]{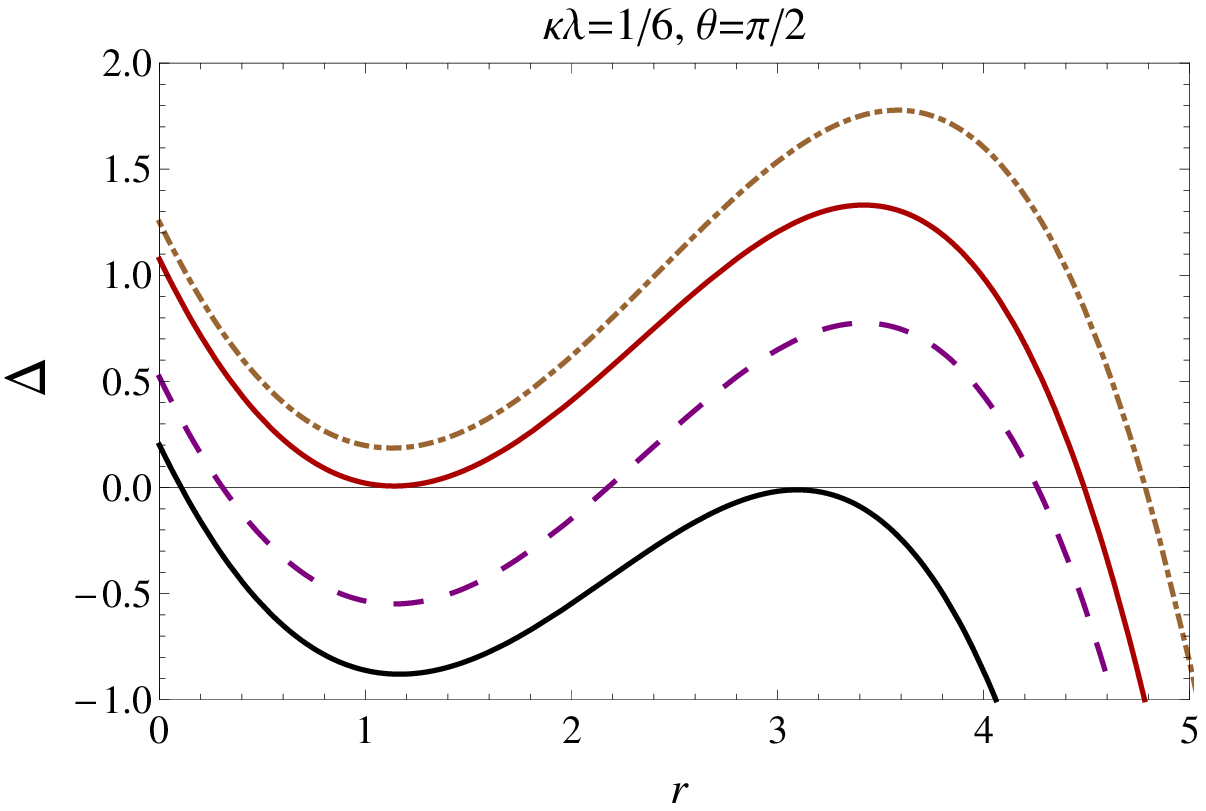}
\caption{Plot showing the variation of $\Delta$ for charged rotating black hole in Rastall theory with $r$ for different values of parameter surrounded by a field of state parameter $\omega_{eff}(\leq-1/3)$. (\textit{Solid black curve}) represent the extremal black hole with inner horizon $r_-$ and outer horizon $r_+=r_q$. (\textit{Dashed purple curve}) represent the black hole with inner horizon $r_-$, outer horizon $r_+$ and cosmological horizon $r_q$. (\textit{Solid red curve}) also represent an another extremal black hole with horizon $r_-=r_+$ and cosmological horizon $r_q$. (\textit{Dot dashed brown curve}) represent the spacetime with naked singularity and cosmological horizon $r_q$.}
\label{Horizonquint1}
\end{figure}
The free quintessential background generates a cosmological horizon of de Sitter type with radius $r_q=1/N_s$. In presence of this field, a Schwarzschild black hole (of radius $2M$ in vacuum) has two horizons   
\beq
r_{-}=\frac{1-\sqrt{1-8MN_s}}{2N_s},\quad r_{+}=\frac{1+\sqrt{1-8MN_s}}{2N_s}
\eeq
provided $(1>8MN_s)$. For small mass black hole, $r_{-}$ approaches the isolated black hole radius and for extremal black hole ($8MN_s=1$) two coincide. An observer which has not fallen into the black hole, and which can still see the black hole despite the inflation, is sandwiched between the two horizons, which is a static.
\begin{table}
 \centering
\begin{tabular}{ c c  c  c  c c }
 \hline
  $\kappa\lambda$     &     $\xi$     & $N_s$ &  $r_-$ &  $r_+$  & $r_q$\\
 \hline
    $\frac{1}{10}$    &      $\frac{-4}{3}$     & +0.05 & 0.431635  & 1.83602 & 7.61504\\
    \hline
    $\frac{1}{8}$ & $\frac{-10}{7}$ & +0.05 & 0.431829  &1.86018 & 6.35694\\
  \hline 
 $ \frac{1}{6}$ & $\frac{-8}{5}$ & +0.05 & 0.432167  & 1.91653 & 4.78257\\
  \hline
   $\frac{2}{5}$ & -3 & -0.05 &  0.4333637 & 0.435003 & 1.33209\\
  \hline
   $\frac{1}{2}$ & -4 & -0.05 &  -  & - & 0.434019\\
  \hline
 $ \frac{-3}{2}$ & $\frac{4}{5}$ & +0.05 & 0.418964  & 1.64076 & -\\
  \hline
   $\frac{5}{2}$ & 4 & -0.05 &  0.583745 & 1.546904 & -\\
    \hline
  \end{tabular}
    \caption{Table showing horizon radii of charged rotating black hole surrounded by quintessence in Rastall theory for various coupling strength. A suitable coupling strength may wipe out the quintessential horizon.}
  \label{quinttable}
\end{table}
The horizon of quintessential KN black hole ($\kappa\lambda=0$) is dictate by (at $\theta=\pi/2$)
\beq
N_s r^3-r^2+2Mr-(a^2+Q^2)=0
\eeq
for parametric values of $M=1$, $a=0.8, Q=0.2$ and $N_s=0.05$, quintessential KN black hole in standard GR has three horizons with radii $r_{-}=0.430792,\quad r_{+}=1.77406,\quad r_q=17.7951.$ A quantitative description of horizon radii for various choices of parameter is shown in Table \ref{quinttable}. It is clear that quintessential horizon radius decrease in the presence of non-zero Rastall coupling, however black hole event horizon radius surprisingly increase. A detailed behavior of $\Delta$ with $r$ is shown in Fig.~\ref{Horizonquint} and Fig.~ \ref{Horizonquint1}. Depending upon the values of parameter ($M, a, Q, N_s$) the number of horizon may vary from three, two or one. However, cosmological quintessential horizon never vanishes unless $N_s\rightarrow 0.$ Therefore, if only one horizon exists, then spacetime describes the quintessential rotating naked singularity. Therefore, a surrounding quintessence field around black hole will behave like a different effective surrounding field whose equation of state parameter depends extensively upon the choice of $\kappa\lambda$.
\section{Thermodynamics}
Wheeler \cite{Ruffini:1971bza, Misner:1974qy} seems to have been the first to notice that a physical system having black hole violates the law of non-decreasing entropy. This necessitate, to assign temperature and entropy to black hole. Having this assertion, we can conclude that a body falling into black hole not only transfer its mass, angular momentum and charge (if any) but its entropy as well. In past 30 years, it has been found that the analogy of black hole physics to thermodynamics is quite far reaching \cite{Gibbons:1977mu, Hawking:1974sw, Bekenstein:1973ur, Bardeen:1973gs, Tharanath:2014uaa}. Thermodynamical quantities associated with black hole depends only upon the geometrical properties of event horizon. 
 
Thermodynamics of black hole in the presence of surrounding field has been widely studied in literature \cite{Chen:2008ra,Sekiwa:2006qj,Gibbons:2004ai,Ghaderi:2016dpi}. Next, we calculate the thermodynamical quantities associated with rotating Rastall black hole  described by Eq. (\ref{metric2}). An extended form of zeroth law of black hole mechanics implies that the surface gravity $\kappa$, the angular velocity $\Omega$ and the electrostatic potential are all locally defined on the horizon and are always constant over the horizon of any stationary black hole.  Bekenstein noticed that one of the property associated with black hole surface (area $A$) resemble the thermodynamical property (entropy), which is a crucial difference between black hole and other thermal system.  Area of the black hole horizon can be calculated by the metric components
\begin{equation}
A_H=\int_{0}^{2\pi}d\phi\int_{0}^{\pi}\sqrt{g_{\theta\theta}g_{\phi\phi}}d\theta,
\end{equation}
It must be noted that the horizons are govern by $\Delta=0$, it is clear from Eq.~(\ref{horizon1}) that horizon radii have a atypical dependency upon $\theta$, however still upon integration we can write the horizon area as
\begin{equation}
A_H=4\pi(r_+^2+a^2),
\end{equation}
where, $r_+$ is outer horizon radius. The universal area law of black hole mechanics gives the entropy of black hole as 
\beq
S=\frac{A_H}{4}=\pi (r_+^2+a^2).
\eeq
In the limiting case of $a=0, Q=0,$ and $N_s=0$ the entropy expression reduces to $S=4\pi M^2$, which is the value for Schwarzschild black hole. Using the null property of Killing vector $\chi^{\mu}=\eta_{t}^{\mu}+\Omega\eta_{\phi}^{\mu}$ at null hypersurface (event horizon), we can calculate the rotational velocity of black hole horizon 
\begin{equation}
\Omega = \frac{\left[\pm\Sigma\sqrt{\Delta}+a\sin\theta(\Sigma-\Delta+a^2\sin^2\theta)\right]}{\left[a^4\sin^4\theta -a^2(-2\Sigma+\Delta)\sin^2\theta+\Sigma^2\right]\sin\theta};
\end{equation}
at horizon $\Delta=0$, we get the horizon rotational frequency reads as
\begin{equation}\label{frequency}
\Omega_H=\frac{a}{r_+^2+a^2}.
\end{equation}
Based on the discussion in previous section, the event horizon radius increase in Rastall theory subject to the condition of positive structure coefficient as compare to that in GR, thus the horizon rotational velocity decrease in the Rastall theory.
Since, black hole behaves as a thermodynamical entity whose temperature $T$ can be calculated from surface gravity  
$\kappa$ evaluated at Killing horizon through
\begin{equation}
\kappa^2=-\frac{1}{2}\chi^{\mu;\nu}\chi_{\mu;\nu}.
\end{equation}
Hawking showed that the black hole temperature is determined by
\begin{equation}
T=\frac{\kappa}{2\pi},\quad \kappa=\frac{\Delta'(r)}{2(r_+^2+a^2)},
\end{equation}
where, $\Delta'(r)$ is spatial derivative of $\Delta(r)$.
In this way, the horizon temperature yields as
\begin{eqnarray}
T&=&\frac{1}{2\pi(r_+^2+a^2)}\left[r_+-M+ \frac{N_sr_+(\xi-2)}{2\Sigma^{\xi/2}}\right],\nn\\
  &=& \frac{1}{4\pi r_+(r_+^2+a^2)}\left[r_+^2-a^2-Q^2+ \frac{N_s}{\Sigma^{\xi/2}}(\Sigma+ r_+^2(\xi-2))\right].
\end{eqnarray}
\begin{figure*}
 \begin{tabular}{c c}
\includegraphics[scale=0.7]{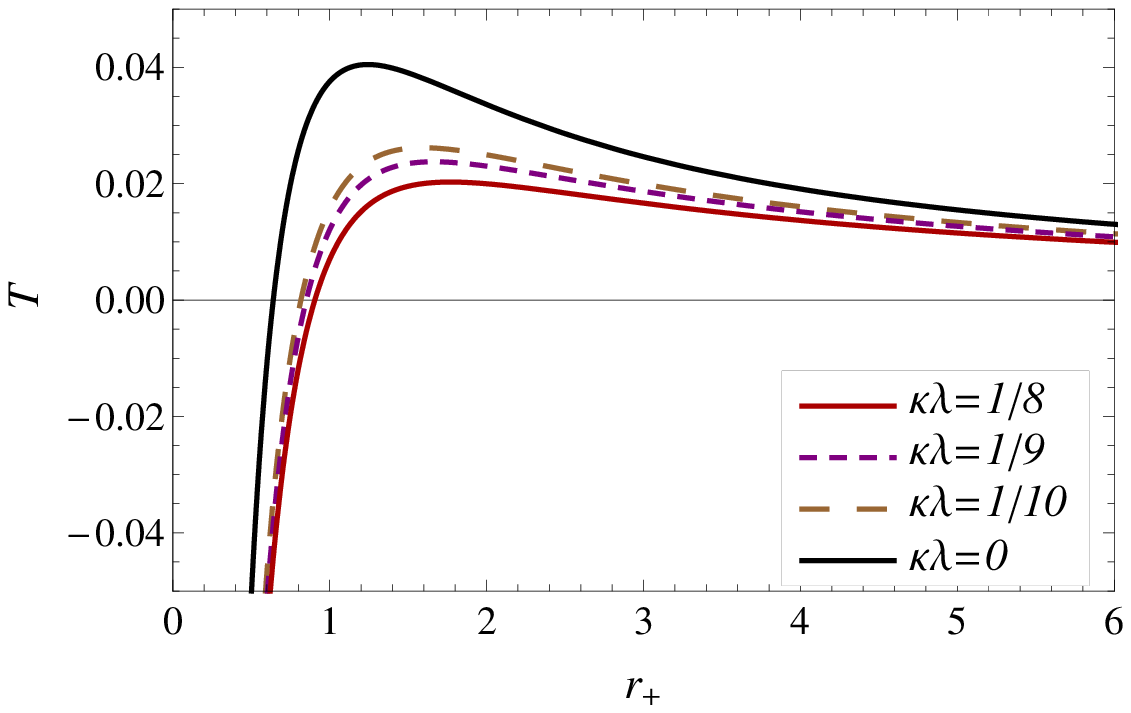}&
\includegraphics[scale=0.7]{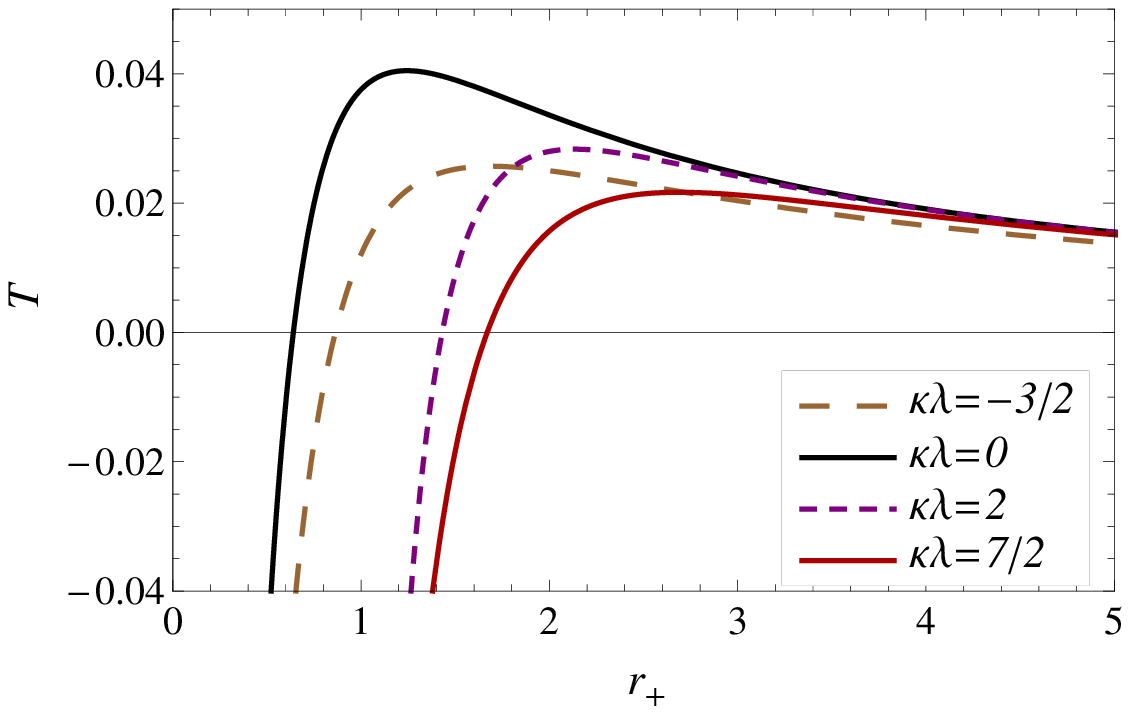} 
\end{tabular}
\caption{Plot showing the variation of Hawking temperature $T$ with horizon radius $r_+$ for different values of Rastall coupling $\kappa\lambda$ for rotating Rastall black hole ($a=0.5, Q=0.1$) surrounded by dust (\textit{Left}) and quintessence field (\textit{Right}). (\textit{Black solid curve}) represent the temperature profile for KN black hole in vacuum.}
\label{BHtemp}
\end{figure*}
This expression leads to the Schwarzschild black hole temperature in the limiting case of $a=0, Q=0, N_s=0$, and electrically charged KN black hole for $N_s=0$ which reads as
\beq
T_{KN}=\frac{(r_+-M)}{2\pi(r_+^2+a^2)}.
\eeq 

From the classical electrodynamics we can calculate the electrostatic potential associated with black hole, which reads as
\beq
\Phi=\frac{Q r_+}{r_+^2+a^2}.
\eeq

Recently, some authors suggest to regard the field structure parameter $N_s$ as a variable and promote it to black hole thermodynamics
\cite{Chen:2008ra, Sekiwa:2006qj}. Therefore, the differential form of the first law of black hole thermodynamics can be written as
\beq
dM=TdS+\Omega dJ+\Phi dQ+\Theta d N_s;\\ \label{entropy}
\eeq
where $\Theta=(dM/dN_s)_{(S,J,Q)}$ is generalized force corresponds to the field variable $N_s$. Furthermore, using this we can calculate the extensive quantity associated with black hole i.e. temperature, angular velocity and electrostatic potential through
\beq
T=\left(\frac{dM}{dS}\right)_{(J,Q,N_s)},\quad
\Omega=\left(\frac{dM}{dJ}\right)_{(S,Q,N_s)},\quad
\Phi=\left(\frac{dM}{dQ}\right)_{(S,J,N_s)}.
\eeq
It is a matter of straightforward calculation to show that these quantity satisfy the first law of black hole thermodynamics. 
 Now, to test the thermodynamical stability of black hole, we need to check the behavior of specific heat of black hole, which is defined as
\begin{eqnarray}
C &=& \frac{dM}{dT}=\left(\frac{dM}{dr_+}\right)\left(\frac{dr_+}{dT}\right),\\
&=& \frac{2\pi(r_+^2+a^2)^2\left((r_+^2-a^2-Q^2)(r_+^2+a^2)^{\xi/2} +N_s(r_+^2(\xi-1)+a^2)  \right)}{(r_+^2+a^2)^{\xi/2}\left(a^2(a^2+Q^2) +(4a^2+3Q^2)r_+^2-r_+^4\right) -N_s\left( a^4+4a^2r_+^2+(\xi^2-1)r_+^4\right)}.\nn
\end{eqnarray}

\begin{figure*}
 \begin{tabular}{c c}
\includegraphics[scale=0.7]{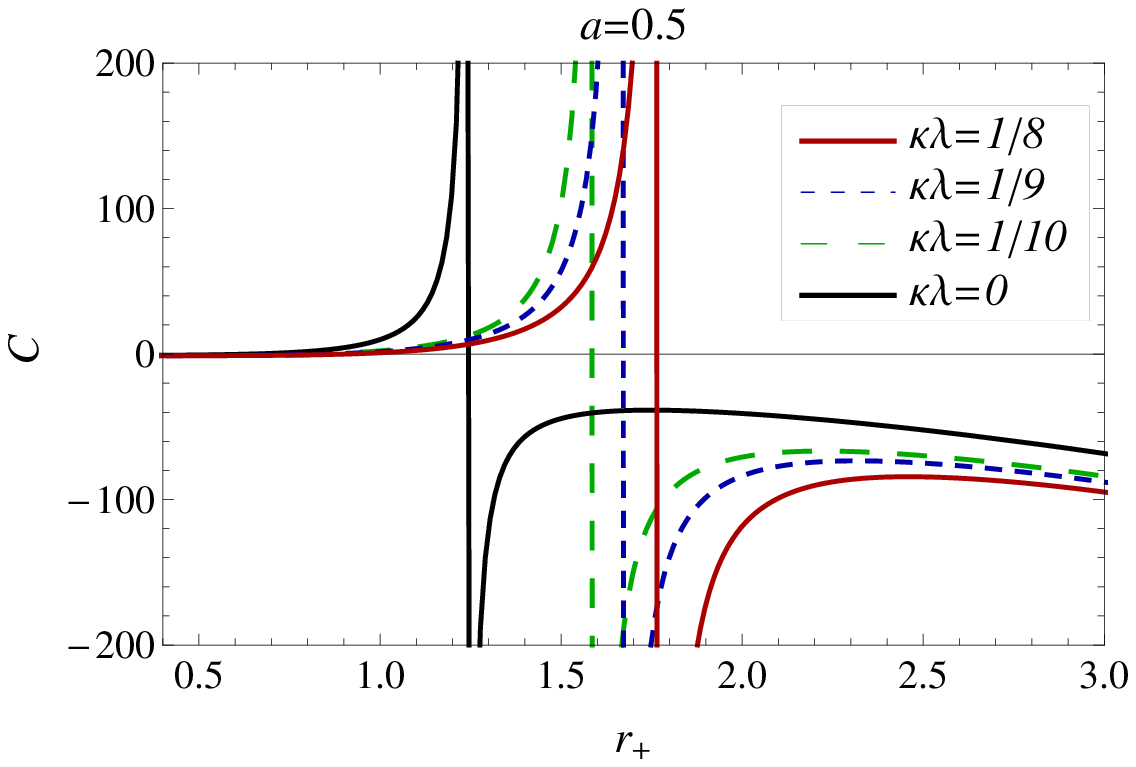}&
\includegraphics[scale=0.7]{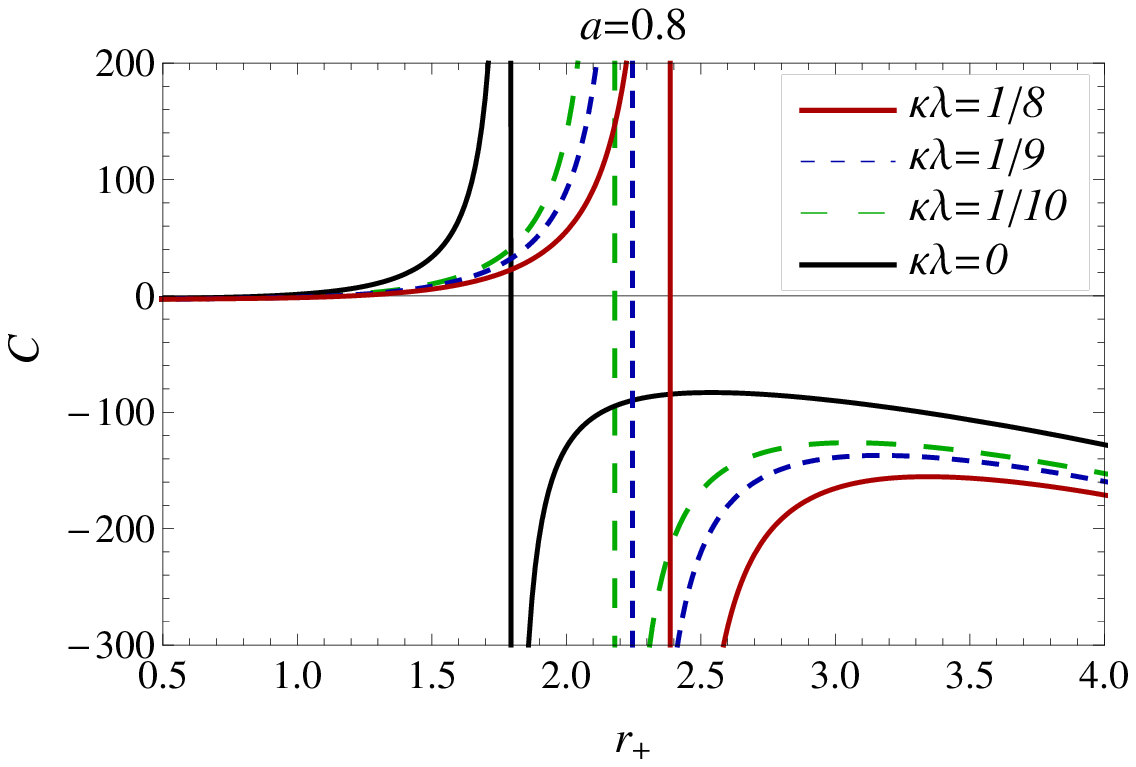} \\
\includegraphics[scale=0.7]{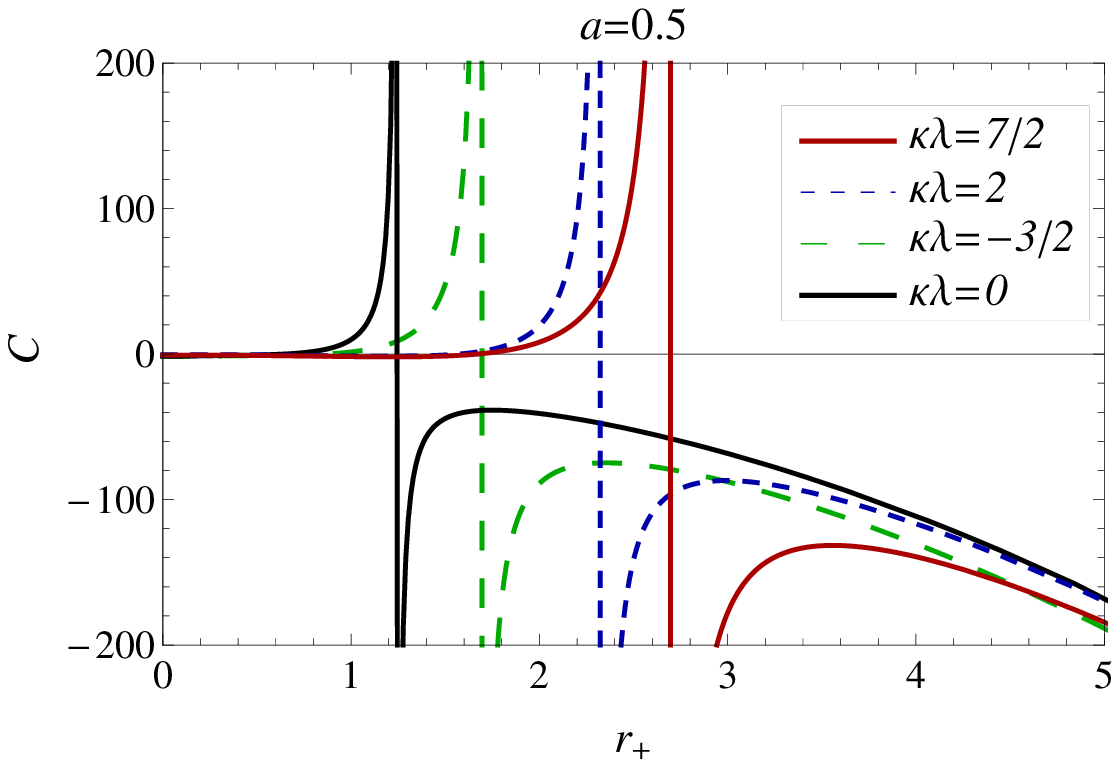}&
\includegraphics[scale=0.7]{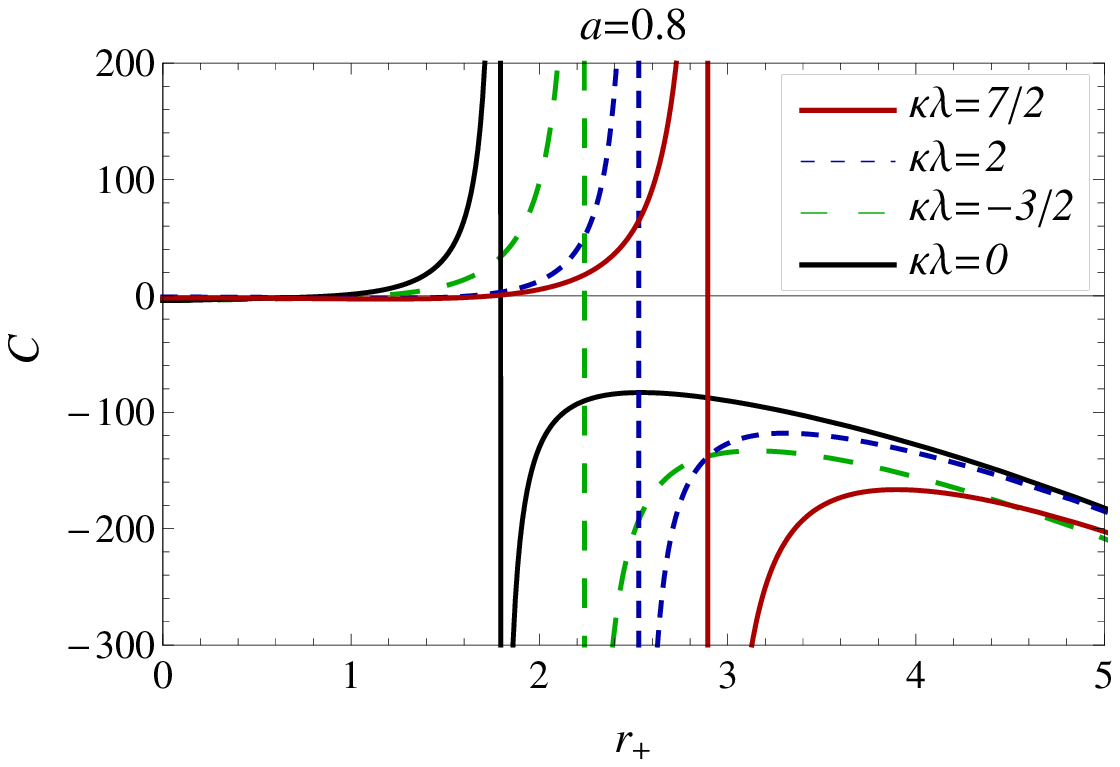}
\end{tabular}
\caption{Heat capacity behavior in terms of horizon radius for rotating Rastall black hole. (\textit{Top}) surrounded by dust and (\textit{Bottom}) surrounded by quintessence field.}
\label{specificheat}
\end{figure*}
The calculated temperature and specific heat are in normalized units. 
In Fig.~\ref{BHtemp} we plotted the temperature profile of black hole surrounded by dust and quintessential matter. During evaporation black hole temperature increase gradually as it shrink, attain a maximum value and then finally falls to absolute zero. It can be easily noticed from Fig.~\ref{BHtemp} that this finite maximum temperature that the black hole can reach depends explicitly upon $\kappa\lambda $.  At zero temperature state black hole evaporation stops completely and a finite sized thermodynamically stable black hole remnant remains. This is well known that black hole local thermodynamical stability is directly related to the sign of specific heat capacity. Indeed, black hole is locally thermodynamical stable only in the region where specific heat is non-negative whereas it is thermodynamical unstable if specific heat is negative. A second order phase transition occur at a point where $C$ is discontinuous and changes it sign abruptly, and it is evident that during evaporation horizon temperature attains its maximum value exactly at the same point (cf. Fig. \ref{specificheat}). From Figs. \ref{BHtemp} and \ref{specificheat}, we can see that rotating Rastall black hole is thermodynamically stable in the region where its horizon temperature decreases with decreasing horizon radius.
\section{The equation of motion and the effective potential}
Particle motion around black hole surrounded by perfect fluid or quintessence has been discuss in various context in literature \cite{Hussain:2016tth, Oteev:2016fbp}.
Here, we would like to investigate the equations of motion of a test particle with rest mass $m_0$ falling into the background of rotating Rastall black hole. To calculate the Center of Mass (CM) energy of the collision on the horizon of the concerned black hole, we must have to derive the 4-velocity, $u^\mu$, of the colliding particle. For the ongoing discussion of the particle motion we shall restrict ourselves to the equatorial plane (i.e. $\theta=\pi/2$, $\Sigma=r^2$).  Since the two Killing vectors of space-time (\ref{metric2}) $\eta_{(t)}^{\mu}$ and $\eta_{(\phi)}^{\mu}$ correspond to time translational and rotational symmetry, therefore we have two conserved quantities associated with them. Let $E$ and $L$ be energy and angular momentum per unit mass of the particle respectively. Thus the defining condition of these conserved quantities are:
\begin{equation}\label{E}
E=g_{\mu\nu} \eta_{(t)}^{\mu} u^\nu=g_{tt} u^t+g_{t\phi}u^\phi,
\end{equation}
\begin{equation}\label{L}
-L=g_{\mu\nu} \eta_{(\phi)}^{\mu} u^\nu=g_{t\phi} u^t+g_{\phi\phi}u^\phi.
\end{equation}
 By solving Eqs. (\ref{E}) and (\ref{L}) simultaneously we can easily calculate the velocity components $u^t$ and $u^{\phi}$, which reads as 
\begin{eqnarray}\label{velocity1}
u^t &=&\frac{1}{\Delta}\left[Er^2+Ea^2\left(2-\frac{\Delta-a^2}{r^2}\right)-La\left(1-\frac{\Delta-a^2}{r^2}\right)\right],\nn\\
u^\phi &=&\frac{1}{\Delta}\left[Ea\left(1-\frac{\Delta-a^2}{r^2}\right)+L\left(\frac{\Delta-a^2}{r^2}\right)\right].
\end{eqnarray}
Furthermore, using the velocity normalization condition {$u^{\mu}u_{\mu}=-1$}, the radial velocity component $u^r$ comes up as
\begin{equation}
u^r=\pm \frac{1}{r^2}\sqrt{E^2\left(a^4+r^4+a^2(2r^2-\Delta)\right)+2aEL(\Delta-r^2-a^2)-L^2(\Delta-a^2)+\Delta r^2}\label{radial}
\end{equation}
 The '$\pm$' sign corresponds to  outgoing and incoming geodesics of the test particles. To calculate the effective potential one must take \cite{Chandraseckar}
\begin{equation}
\frac{1}{2}\left(u^r\right){^2}+V_{eff}=E^2
\end{equation}

\begin{figure*}
 \begin{tabular}{c c}
\includegraphics[scale=0.7]{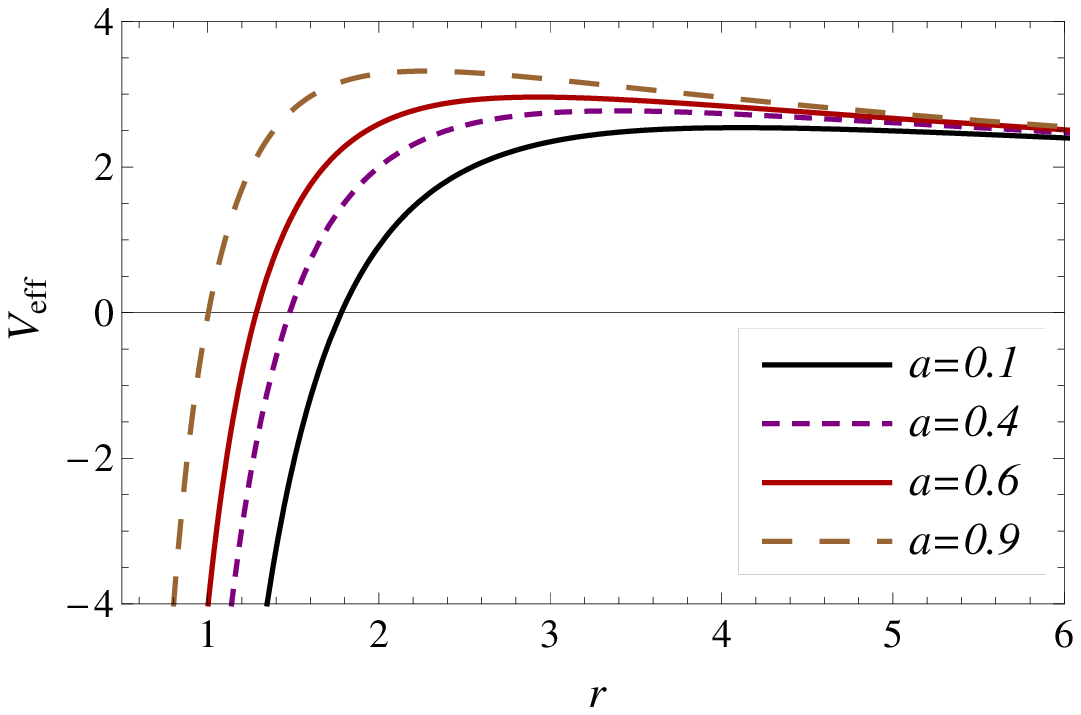}&
\includegraphics[scale=0.7]{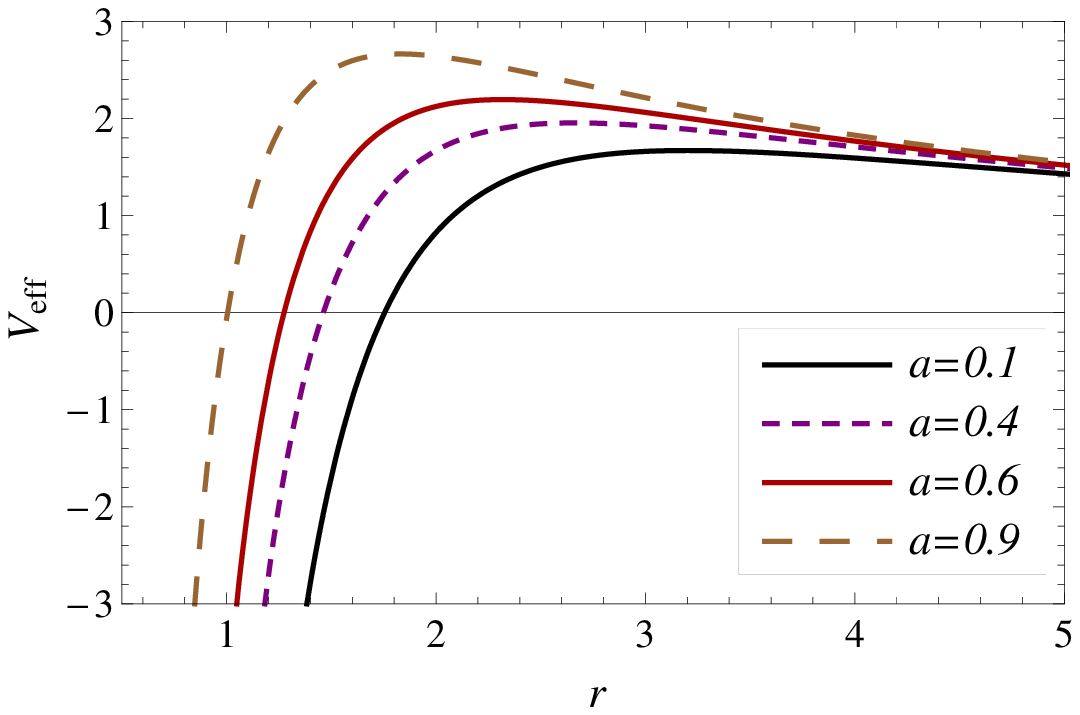} 
\end{tabular}
\caption{Plot showing the variation of effective potential with $r$ for different values of parameter $a$  for charged rotating black hole ($Q=0.3$). (\textit{Left}) Surrounded by dust field ($\kappa\lambda=1/8$) and (\textit{Right}) surrounded by quintessence field ($\kappa\lambda=-3/2$).}
\label{potential}
\end{figure*}

Therefore, the form of the effective potential looks like
\begin{equation}
V_{eff}(r)=-\frac{1}{2r^4}\left[E^2\left(a^4+r^4+a^2(2r^2-\Delta)\right)+2aEL(\Delta-r^2-a^2)-L^2(\Delta-a^2)+\Delta r^2\right]+E^2.
\end{equation}
In Fig.~\ref{potential} we compare the effective potential of test particle in vicinity of black hole surrounded by dust and quintessence. 
Penrose \cite{rp} in 1969 suggested that energy can be extracted from a rotating black hole. Rotating black hole has a unique surface called  ergosphere(SLS), which plays a significant role in turning time-like Killing vector into space-like inside its boundary. The region between event horizon and SLS is called ergoregion, which is the key for extracting energy from black hole. The SLS for metric (\ref{metric2}) can be determined from the null property of $\eta_{(t)}$ at SLS
\bea
(\eta_{(t)})^{\mu}(\eta_{(t)})_{\mu}=g_{tt}=0\nn \\
1-\frac{(2M r-Q^2)}{\Sigma}-\frac{N_s}{\Sigma^{(\xi-4)/2}}=0\label{SLS}
\eea
Two physical solutions of Eq.~(\ref{SLS}) are $r_{SLS}^+$ and $r_{SLS}^-$, where $r_{SLS}^+>r_{SLS}^-$.
Region between $r_+$ and $r_{SLS}^+$ called ergoregion. Unlike in the KN metric, the ergoregion in this metric (\ref{metric2}) depends upon $\kappa\lambda$. 
Using Eqs.~(\ref{velocity1}), (\ref{radial}) and the normalization property of test particle four velocity in ergoregion, we can write the equation of motion as:
\beq
\alpha E^2-2\beta E+\gamma=0\label{Eequation}
\eeq
with
\beq
 \alpha=r^4+a^4-a^2(\Delta-2r^2),\quad \beta=La(r^2+a^2-\Delta),\quad \gamma=r^2\Delta-L^2(\Delta-a^2)-r^4(u^r)^2.\\ \label{Eequation2}
\eeq
Inside SLS, $g_{tt}$ change its sign and hence it is possible that energy of particle defined from $E=-g_{tt}\eta^{t}u^t$ becomes negative as perceived by the observer at spatial infinity. In Penrose energy extraction process, a particle splits into two particles in ergoregion, one with negative energy which falls into the black hole (called injected particle), while another with positive energy comes out of ergosurface. Absorption of negative energy particle by rotating black hole is the only requirement for energy extraction. From  Eqs.~(\ref{Eequation}) and (\ref{Eequation2}), it is clear that for negative energy state ($E=(\beta+\sqrt{\beta^2+\alpha\gamma})/\alpha$), we must have $\alpha>0$, $\beta<0$ and $\gamma>0$, which as a result turns out as a condition $La<0$. Thus, in order to extract energy from rotating black hole we must have a injecting particle with $L<0$, which simply means the retrograde motion in ergoregion.
\section{Conclusion}
 The Rastall theory \cite{Rastall:1973nw, Rastall:1976uh} questions the validity of covariant conservation of energy-momentum tensor atleast in the curved spacetime, and proposed that covariant derivative of stress tensor depends upon the gradient of Ricci scalar. This theory is in full agreement with the standard GR in the limiting case of vanishing coupling between  matter field and geometry. As a result of Rastall coupling, a field described by stress tensor $T_{\mu\nu}$ in GR behave differently as $T'_{\mu\nu}$ and so does with different state parameter in Rastall theory, i.e. a fluid respecting the strong energy condition in GR may violate it in Rastall theory and can account for accelerating expansion of universe. Interestingly, a single surrounding quintessential field may regulate either the expansion or contraction depending upon the choice of Rastall coupling parameter.  In this paper, we obtained and analyze the solution of Einstein equation for stationary axially symmetric charged\-\ rotating black hole surrounded by perfect fluid in the context of Rastall theory, i.e., rotating Rastall black hole.  We extensively discussed the case of dust and quintessence surrounding the rotating Rastall black hole. In the limit of vanishing Rastall coupling ($\lambda\rightarrow 0$) we obtain the rotating Kiselev like black hole in GR. By comparing the solution in Rastall theory with that in standard GR, we realized that effective state parameter depends upon the  values of $\kappa\lambda$. For different values of state parameter $\omega_s$ we study the horizon structure. Non-minimal coupling between gravity and surrounding matter field modified the black hole horizon properties. The spacetime geometry in Rastall theory is different from that in standard GR. In the case of quintessential black hole apart from the black hole horizon an extra horizon is also present in spacetime. Depending upon the numerical value of black hole parameter $M, a, Q$ and Rastall coupling $\lambda$ of surrounding fluid characterize by structure parameter $N_s$ the number of horizon in spacetime may vary from three, two or one. However, in such field cosmological horizon will never vanish unless surrounding field itself does not fade away. Various possibilities for extremal black hole are also studied. It is noted that quintessential cosmological horizon radius surprisingly decrease in the Rastall theory, however event horizon radius increase as compare to that in standard GR, unless the extremal condition reach, subject to the positive structure parameter. Interestingly, the charged\-\ rotating black hole in Rastall theory also respect the first law of thermodynamics. Explicit expression for horizon temperature, entropy and specific heat are also obtained, which in the limits of $\kappa\lambda\rightarrow 0$ matches with that of Kiselev\-\ like black hole. It can be noted that horizon temperature declined in the presence of non-zero Rastall coupling.
\section{Acknowledgements}
S.G.G. would like to thanks SERB-DST Research Project Grant No. SB/S2/HEP-008/2014 and DST INDO-SA bilateral project DST/INT/South Africa/P-06/2016 and also to IUCAA, Pune for the hospitality while this work was being done. R.K. would like to thanks UGC for providing Junior Research Fellowship.

\section{Added in proff} After this work was completed, we learned of a similar work by Zhaoyi Xu et al. \cite{Xu:2017bix}, which appeared in arXiv a couple of days before.

\noindent
\end{document}